\pdfoutput=1

\documentclass{vldb}

\usepackage{color}
\usepackage{xspace}
\usepackage[normalem]{ulem}
\usepackage{url}
\usepackage{epsfig}
\usepackage{subfigure}
\usepackage{multirow}
\usepackage{amssymb}
\usepackage{cite}
\usepackage{listings}
\usepackage{epstopdf}
\usepackage{balance}

\usepackage{algorithm}
\usepackage{algpseudocode}
\algnewcommand\algorithmicinput{\textbf{Input:}}
\algnewcommand\INPUT{\item[\algorithmicinput]}
\algnewcommand\algorithmicoutput{\textbf{Output:}}
\algnewcommand\OUTPUT{\item[\algorithmicoutput]}

\newcommand{\eat}[1]{}
\newcommand{\bird}{{STEP}\xspace}
\newcommand{\bl}{{BDL}\xspace}
\newcommand{\blib}{{libBirdee}\xspace}

\newcommand{\execute}{{\sf .bde}\xspace}

\vldbTitle{\bird : A Distributed Multi-threading Framework Towards Efficient Data Analytics}
\vldbAuthors{Yijie Mei, Yanyan Shen, Yanmin Zhu, Linpeng Huang}
\vldbDOI{https://doi.org/TBD}
\vldbVolume{12}
\vldbNumber{xxx}
\vldbYear{2019}

\makeatletter
\def\@copyrightspace{\relax}
\makeatother

\begin{document}


\title{\bird : A Distributed Multi-threading Framework Towards Efficient Data Analytics}



%
%
%
%


\author{Yijie Mei, Yanyan Shen, Yanmin Zhu, Linpeng Huang \\
	Department of Computer Science and Engineering\\
	Shanghai Jiao Tong University \\
	\{myjisgreat, shenyy, yzhu, lphuang\}@sjtu.edu.cn
}

\maketitle

\begin{abstract}
Various general-purpose distributed systems have been proposed to cope with high-diversity applications in the pipeline of Big Data analytics. Most of them provide simple yet effective primitives to simplify distributed programming. While the rigid primitives offer great ease of use to savvy programmers, they probably compromise efficiency in performance and flexibility in data representation and programming specifications, which are critical properties in real systems. In this paper, we discuss the limitations of coarse-grained primitives and aim to provide an alternative for users to have flexible control over distributed programs and operate globally shared data more efficiently. We develop STEP, a novel distributed framework based on in-memory key-value store. The key idea of STEP is to adapt multi-threading in a single machine to a distributed environment. STEP enables users to take fine-grained control over distributed threads and apply task-specific optimizations in a flexible manner. The underlying key-value store serves as distributed shared memory to keep globally shared data. To ensure ease-of-use, STEP offers plentiful effective interfaces in terms of distributed shared data manipulation, cluster management, distributed thread management and synchronization. We conduct extensive experimental studies to evaluate the performance of STEP using real data sets. The results show that STEP outperforms the state-of-the-art general-purpose distributed  systems as well as a specialized ML platform in many real applications.
\end{abstract}

\section{Introduction}\label{sec:introduction}
Big Data analytics is broadly defined to be a pipeline involving several distinct phases, from data acquisition and cleaning to data integration, and finally data modeling and interpretation~\cite{whitepaper}. This pipeline poses various challenges in developing distributed systems for Big Data analytics. In particular, the system should fulfill multiple design goals: (1) be \emph{flexible} to cope with high-diversity applications in the pipeline; (2) be \emph{efficient and scalable} to handle ever-increasing data; (3) provide \emph{easy-to-use} APIs to shorten the learning curve for programmers; and (4) be \emph{insensitive} to underlying operation systems to facilitate easy deployment. 

Numerous efforts have been devoted to addressing this ``multi-objective optimization'' problem, which consequently lead to a proliferation of \emph{general-purpose} distributed systems for Big Data analytics over the last few decades. Existing general-purpose distributed frameworks typically fall into two categories: (1) \emph{disk-based} solutions including MapReduce~\cite{mapreduce} (and its open-source implementation Hadoop~\cite{hadoop}) and Dryad~\cite{dryad}; (2) \emph{memory-based} solutions such as Spark~\cite{spark}.
Spark introduces an in-memory computation model to eliminate expensive I/O cost for iterative computation tasks and is reported to be over 10x faster than Hadoop~\cite{spark}. 

The \emph{generalization} ability of existing distributed systems comes from their simple yet effective functional programming primitives, e.g., \emph{map} and \emph{reduce} functions in Hadoop, \emph{transformation} and \emph{action} operations over immutable data abstraction -- RDDs in Spark. 
While these high-level operations simplify parallel programming by separating programming paradigm from implementation details, they are still limited in expressiveness and functionality~\cite{piccolo,husky}.
More importantly, the adoption of fixed primitives potentially drive the development of systems towards ease-of-use extreme with compromise of other objectives. 


First of all, high-level operations inhibit the opportunities of task-specific optimizations. Particularly, compared with specialized distributed systems, many general-purpose systems perform poorly in real applications. 
For instance, specialized graph processing systems such as PowerGraph~\cite{powergraph} and Giraph~\cite{giraph} can be 10 times faster than Spark and its graph analytics extension GraphX~\cite{graphx, husky}; and the distributed machine learning platform Petuum~\cite{petuum} outperforms MLlib~\cite{mllib} (ML library based on Spark) significantly for K-means clustering (in our experiment), thanks to its soft synchronization mechanism 
and the core component parameter server for global parameter management.


Second,  enforcing rigid interfaces limits the flexibility in data representation and programming specification. Real-life data sets may come in various types: structured, semi-structured and unstructured with latent semantics encoded. Transforming these data into predefined forms such as RDDs or key-value pairs can be time-consuming and even cause information loss. Furthermore, specific programming models implemented beneath the interfaces does not allow fine-grained user access control. For example, data processing in Spark are performed via a sequence of RDD transformations while a MapReduce job consists of a map phase followed by a reduce phase. Designing algorithms that fit into these programming models is often a non-trivial task. This can be verified by the emergence of several Spark extensions, e.g., GraphX and MLlib that enable natural expressions for graph analytics and machine learning tasks, respectively.

Finally, we find most provided operations are insufficient to manage globally shared data. 
Note that Spark does not provide explicit global data management and data sharing is performed inefficiently via broadcasting.
However, managing shared data is inevitably important in most machine learning applications, where the model (e.g., coefficients for regression analysis) to be trained is globally accessed and updated by a cluster of compute nodes iteratively. Things become even worse in the presence of Big Model training, e.g., deep learning models with billions of parameters~\cite{dl} and topic models with up to hundreds of topics~\cite{lightlda}.

To this end, we propose a novel general-purpose distributed framework, named \textbf{\bird}, towards flexible and efficient data analytics. 
The key idea of \bird is to adapt multi-threading in a single machine to a distributed environment, which is inspired by two important observations: 1) very often, users feel more comfortable writing their programs in procedural languages without fixed primitives; 
2) multi-threaded programming such as Pthreads~\cite{pthread} has achieved great success in expressing various programming models and accelerating task execution in a single machine.
Hence, rather than define primitive operations, \bird allows users to write distributed multi-threaded programs in a similar way as they accelerate task execution with multi-threading in a centralized manner.
We believe that \bird is a suitable alternative for users who would like to have more control over distributed systems to deliver efficient programs.

With \bird, users can specify the number of threads to run over each compute node in a cluster. Furthermore, they are allowed to take fine-grained control over threads' behaviors via a user-defined {\sf thread\_proc} function, where object-oriented programming and various task-specific optimizations can be easily performed.
Unlike multi-threading in a single machine, distributed threads running in different compute nodes do not have shared memory space. \bird addresses this problem by leveraging off-the-shelf in-memory key-value stores so that data maintained in key-value stores are globally accessible to all threads in \bird cluster.  

The implementation of \bird is challenging due to the following two reasons.
First and foremost, the flexibility and expressiveness of distributed multi-threading should not increase the difficulty of programming.
High-level interfaces for remote thread management and communication are strongly demanded to make the system easy to use.
Second, shared data manipulation requires user programs to interact with the underlying key-value store frequently. However, the dependency on specific key-value store implementation may easily make user programs less reusable. 

To address the challenges, \bird offers plentiful interfaces in terms of cluster management, distributed thread management and synchronization, which hide complex implementation details of distributed multi-threading. 
Most interfaces are designed to be ``aligned'' with Pthreads APIs so that programmers who are familiar with Pthreads programming can write distributed multi-threaded \bird programs effectively.
Furthermore, \bird provides abstraction layers to decouple user programs with specific key-value store implementation and supplies users with easy-to-use distributed shared data manipulation interfaces in C++. {By using these interfaces, manipulating shared data in \bird can be expressed as simply as operating local variables. For example, programmers can use normal operator ``='' for shared variable assignment.}
All these efforts differentiate \bird system from MPI-based data analytics solutions~\cite{mpiomp,mpitensorflow,trilinos} which have some deficiencies in programming, e.g., users are responsible for handling low-level message passing by hand and exploiting globally shared data via intra-node data transfer.

To summarize, the main contributions of our work include:

$\bullet$ We propose a novel general-purpose distributed framework named \bird, which facilitates efficient multi-threaded programming in a distributed manner. With \bird, users can perform fine-grained control over distributed programs and enforce various task-specific optimizations. 

$\bullet$ We develop various easy-to-use interfaces in \bird for distributed shared data manipulation, cluster and distributed thread management, which enable users to deliver distributed multi-threaded \bird programs more effectively.


$\bullet$ \bird leverages key-value stores to maintain shared data among distributed threads. We provide abstraction layers to separate shared data access with specific key-value store implementation. An additional cache layer is used to alleviate the heavy burden on key-value store throughput.

$\bullet$ We conduct extensive experiments to evaluate the performance of \bird using various applications and real data sets. The experimental results show that \bird outperforms Spark on logistic regression, K-means and NMF by 8.6-29x, and runs up to 5.4x faster than specialized ML platform Petuum on K-means and NMF, and up to 3.4x faster than general purposed distributed platform Husky on PageRank.

The remainder of the paper is organized as follows.
Section~\ref{background} contains some background for \bird system. 
Section~\ref{ov} provides an overview of \bird framework. 
Section~\ref{program} introduces key interfaces of \bird.
The implementation details are described in Section~\ref{impl}. 
We evaluate our system in Section~\ref{sec:exp} and discuss related work in Section~\ref{relwork}. 
Finally, we conclude our paper in Section~\ref{sec:conclusion}.

\section{Background} \label{background}

\begin{table}[t]
	\centering
	\caption{\small Support from key-value stores}\label{tab:kv}
	\scriptsize
	\begin{tabular}{|l|l|}  \hline
		\textbf{APIs} & \textbf{Definitions}\\ \hline
		Get & Getting the value of a key\\ \hline
		Set & Setting the value of a key\\	\hline
		MGet & Getting values of multiple keys\\	\hline
		Insert & Atomic key-value pair insertion \\\hline
		Inc/Dec & Atomic increment/decrement on Integers \\\hline
		Delete & Delete an existing key and its value\\\hline		
	\end{tabular}
	\vspace{-.2in}
\end{table}



{\bf Pthreads Programming.}
POSIX threads (Pthreads)~\cite{pthread} is an implementation of the standardized C language multi-threading programming interface introduced by IEEE POSIX 1003.1c standard. It can be found on almost any modern POSIX-compliant OS and is ideally suited for parallel programming where multiple threads work concurrently. Pthreads APIs allow developers to manage the life cycle of threads (e.g., creation, joining, terminating) and express various programming models to maximize thread-level parallelism, e.g., boss-worker, pipeline, peer.
%
Every thread has access to a globally shared memory as well as its own private memory space. Pthreads library also includes APIs to synchronize the accesses to shared data and coordinate the activities among threads e.g., semaphore, barrier. 
We refer readers to more details in~\cite{pthread}.

Pthreads library gains great popularity due to its flexible programming models and the light-weight shared memory management. This inspires us to believe that Pthreads programming has promising potential in flexible and efficient distributed computing for Big Data analytics. However, implementing Pthreads APIs in a distributed environment (especially the shared-nothing architecture) is not merely an extension of the centralized version, and hence we propose \bird as an end-to-end solution to this challenge. 
{\bf Key-value Store Support.}
Distributed in-memory key-value stores such as Redis\footnote{\small\url{http://redis.io/}}, Memcached~\cite{memcached},  MICA~\cite{mica} and HyperDex~\cite{hyperdex} typically act as distributed hashtables to support fast access of object values given unique object keys. The keys and values of objects can have different sizes and data types (i.e., primitive or composite types).
Various hashing techniques are proposed to enable efficient key lookups for object insertion, retrieval and update. 

\bird uses distributed in-memory key-value stores to manage globally shared data for all threads in \bird cluster. To do this, we require underlying key-value store to provide several shared data operations, as shown in Table~\ref{tab:kv}.
While each individual key-value store provides slightly different interfaces for query answering, we observe that most existing key-value stores meet our requirement. Note that \bird introduces high-level distributed shared data manipulation interfaces (above the actual interfaces from key-value stores) so that developers can operate globally shared data without invoking any interfaces from particular key-value stores. 

\section{The \bird Framework} \label{ov}

\begin{figure}[t]
	\centering
	\includegraphics[height=3.8cm,width=0.95\linewidth]{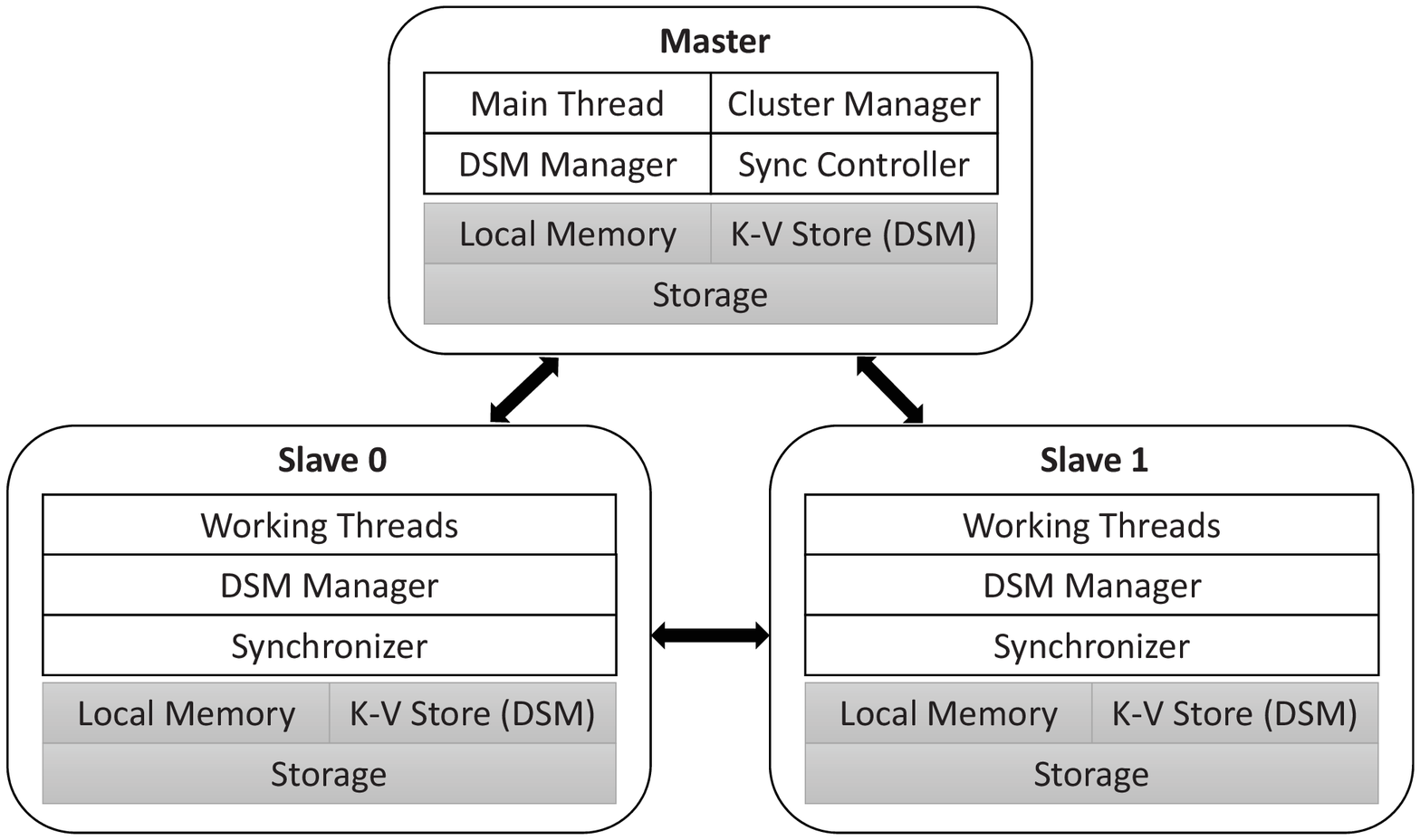}
	\vspace{-.1in}
	\caption{\small \bird architecture}
	\label{fig:ov}
	\vspace{-.2in}
\end{figure}

This section introduces the \bird system and describes how distributed multi-threading is performed in \bird.

We show the architecture of \bird in Figure~\ref{fig:ov}. \bird system is deployed on a cluster of well-connected compute nodes together with a distributed in-memory key-value store. It employs the master-slave architecture, where one node in the cluster is selected as the master and the others are slaves. 
\bird master and slaves have their own local memory space and manage their own threads. They use the in-memory key-value store as distributed shared memory (DSM). That is, data maintained in the key-value store is accessible to all threads running in \bird cluster. 
At a high level, distributed shared memory (DSM) ties all threads together which is similar to centralized shared memory used by multi-threaded programs in a single node.

\bird master plays four roles: \emph{main thread}, \emph{cluster manager}, \emph{DSM manager}, and \emph{sync controller}.
Specifically, the \emph{main thread} is the entry point of a submitted \bird job. Upon creation, the main thread decides the set of slaves and the number of working threads in each slave. It can also declare globally shared data (stored in distributed key-value store) that is accessible to all threads, and specify the behaviors of threads, e.g., when to reach a global synchronization point.
The \emph{cluster manager} is responsible for setting up the cluster and establishing communication channels among \bird master and slaves during initialization. 
%
The \emph{DSM manager} initializes the distributed in-memory key-value store and broadcasts store information (e.g., IP addresses and ports of key-value store servers) to the slaves.
%
The \emph{sync controller} coordinates all the slaves via message passing. It forwards synchronization messages to the slaves and collects responding messages. It also decides when to resume the execution of blocked working threads in slaves.

Every \bird slave manages a couple of \emph{working threads} that execute user-defined program over a subset of data concurrently. Working threads in the same slave leverage multiprocessor to achieve thread-level parallelism while those in different slaves perform distributed computing and communicate with each other via shared data in DSM or network messages. Therefore, similar to \bird master, every \bird slave consists of a \emph{DSM manager} to access shared data in distributed key-value store and a \emph{synchronizer} that is responsible for processing synchronization messages from master to block/unblock working threads accordingly. 



{\textbf{Execution overview.}} 
The execution of a \bird job consists of three phases, \emph{initialization}, followed by \emph{distributed multi-threaded execution} and finally an \emph{output} phase. 

$\bullet$ Phase 1: During initialization, \bird slaves are first started, waiting for the connection from the master. \bird master sets up the cluster and the distributed key-value store. It then establishes connections with selected slaves. Master's main thread then declares globally shared data in DSM and creates working threads on each slave via \bird interfaces for distributed thread management. After that, it notifies slaves of the \emph{entry function} (i.e., {\sf thread\_proc} function) to be executed by slaves' local working threads. 

$\bullet$ Phase 2: Working threads in \bird slaves start their execution by invoking the entry function in parallel with each other. Note that
different working threads may share the same entry function that is user-defined. To express different computation logic for working threads, we assign an identifier to each working thread and allow entry function to use thread identifier as one of its input arguments.
Typically, an entry function is designed to process a subset of input data, operate shared data in DSM and communicate with other threads (main or working threads) based on \bird interfaces. As we adapt the idea of Pthreads programming to \bird system, various computation and communication patterns among threads can be flexibly expressed by users (similar to multi-threaded programming in a single node). 
Upon completion of the entry function, all threads can send synchronization messages to master's sync controller to perform a global synchronization. 

$\bullet$ Phase 3: The output of a \bird job is application-dependent. For machine learning tasks, the output is typically the computed model parameters that reside in DSM for global access. For graph analytics tasks such as PageRank computation, each working thread performs PageRank computation over a subset of vertices iteratively and outputs the resulting PageRank values to its local storage. In either case, the main thread in \bird master can optionally collect output in DSM or slaves' storage to get the complete final results.


\bird allows users to get full control over distributed threads in master and slaves, including the computation logics and communication patterns. However, programming difficulty is increased as a side effect. In particular, this difficulty arises from two aspects: i) complex interaction with underlying key-value store for shared data manipulation; ii) fine-grained cluster and thread management in a distributed manner.
In what follows, we introduce easy-to-use \bird interfaces to hide complex low-level details and enable users to deliver distributed multi-threaded programs with \bird more efficiently.


\section{Programming Interfaces} \label{program}

\bird provides effective programming interfaces to simplify DSM data operations, cluster and distributed thread management, synchronization and vector accumulation.
We illustrate \bird programming interfaces in C++ language.


\subsection{DSM Data Declaration and Manipulation}\label{sec:bl}

We consider three kinds of shared data, \emph{shared variables, shared objects and shared arrays}, to be store in DSM. 
Shared data can be declared in a primitive type (i.e., \texttt{int}, \texttt{float}, \texttt{double}, etc) or a reference type pointing to a shared object or shared array.

{\bf Shared variable.} 
\bird allows developers to use the macro \texttt{DefGlobal(NAME,TYPE)} to define a \emph{shared variable} in DSM, where \texttt{NAME} is the variable name and \texttt{TYPE} is the variable type. After declaration, developers can manipulate shared variables in the same way as normal local variables. For example, they can assign a new value to a shared variable by operator ``='', and the data stored in DSM will be updated accordingly.

{\bf Shared object.}
\bird supports key object-oriented features for shared classes in DSM, including templates, dynamic dispatch for virtual functions, encapsulation, inheritance and polymorphism.
We introduce a base class \texttt{DObject} in \bird and developers should extend \texttt{DObject} or a sub-class of it to construct their own \emph{shared classes}. Similar to shared variables, developers can use macro \texttt{Def(NAME,TYPE)} in the class body to declare the member variable of a shared class.
%
%
%
In \bird, all instances (i.e., objects) of a shared class are \emph{shared objects} that will be stored in DSM.
We provide two APIs \texttt{NewObj} and \texttt{DelObj} to create and delete a shared object, respectively. 
\texttt{NewObj} function returns a reference \texttt{Ref<Class\_Name>} to the newly created shared object in DSM which behaves like a normal pointer in C++. The members of a shared object can be accessed by the ''-\textgreater'' operator through the object reference.


\eat{The following C++ example allocates, accesses and deletes an ``LinkedNode'' object in DSM. Note that the definition for shared class ``LinkedNode'' is omitted for simplicity.
	\begin{lstlisting}
	Ref<LinkedNode> ptr1 = NewObj<LinkedNode>();
	ptr1->next = ptr1;
	DelObj(ptr1);
	\end{lstlisting}}

{\bf Shared array.}
%
\bird provides \texttt{NewArray} and \texttt{DelArray} APIs to allocate and deallocate a shared array in DSM, respectively.
References to shared arrays are defined by a template class \texttt{Array<TYPE>}.
Developers can use indexing operator ``[ ]'' to access elements in a shared array, as normal arrays in C++.
Moreover, \bird allows developers to perform batch operations over arrays. For example, \texttt{CopyTo} function copies a shared array in DSM to a local array, and \texttt{CopyFrom} function copies data in the opposite direction.
Below is an example showing how to operate a shared array \texttt{arr} in \bird. 
\\
\begin{lstlisting}[caption= Shared array example in \bird]
Array<float> arr = NewArray<float>(10); //shared array
arr[4] = 3.14;
float local_buf[3] = {1,2,3}; //local array
arr->CopyFrom(local_buf, 0, 3);
DelArray(arr);
\end{lstlisting}

\eat{
	\subsection{\bird Programming Interface}\label{sec:interface}
	
	We now introduce effective interfaces in terms of cluster and remote thread management. These interfaces enable efficient distributed multi-threaded programming with \bird. Note that the distributed programming APIs for \bl and \blib are essentially the same. In this section, we discuss the interfaces for \bl for simplicity.
}

\subsection{Cluster and Distributed Thread Management}\label{sec:api:cluster}

List~\ref{code:api1} shows the main interfaces for cluster and distributed thread management in \bird.\\ 
\begin{lstlisting}[caption= Main APIs for cluster and distributed thread management, label=code:api1]
extern void HelperInitCluster(int argc,char* argv[]);
extern void CloseCluster();
	class DThread : public DObject{
	ThreadState GetState();
	DThread(thread_proc func, int node_id, uint32_t param);	
};
\end{lstlisting}

\noindent {\bf Cluster management.}
The \texttt{HelperInitCluster} API is responsible for initializing \bird environment and establishing connections among the compute nodes during initialization. 
This function acts differently on the master and slaves. 
Specifically, it parses the arguments from the command line, and then decides whether the current process will run under master mode or slave mode.
In master mode, \texttt{HelperInitCluster} initializes the cluster by i) reading the settings from configuration file, ii) connecting \bird master to the selected slaves and key-value store servers, and iii) forwarding configuration information to all slaves. In slave mode, \texttt{HelperInitCluster} makes the slave node wait for the connection from the master and respond to the master's requests.


The \texttt{CloseCluster} API is used to shut down the cluster, which can only be invoked by \bird master.

\noindent {\bf Thread management.}
\bird allows users to specify the number of working threads to be created in each slave. This is achieved by using the \texttt{DThread} class, a pre-defined shared class in \bird. To declare a working thread on a slave node, users can create a \texttt{DThread} object using \texttt{NewObj} API.


The constructor of \texttt{DThread} takes three arguments: i) {\sf func} is a user-defined \emph{entry function} (i.e., {\sf thread\_proc} function) for working threads; ii) {\sf node\_id} is the ID of the slave node where the working thread is created; iii) {\sf param} is the parameter forwarded to the user-defined \emph{entry function}. Users can get the state of a thread (i.e., alive or completed) via the member function \texttt{GetState}. 
Note that we declare \texttt{DThread} as a shared class so that all its instances are stored in DSM and publicly available to \bird cluster, which is critical for communication among \bird master and slaves.

\subsection{Distributed Thread Synchronization}\label{sec:api:sync}

List~\ref{code:api2} lists important \bird interfaces for synchronizing distributed threads. Both \texttt{DBarrier} and \texttt{DSemaphore} are encapsulated as shared classes (i.e., inherited from \texttt{DObject}) whose instances are accessible to all threads in \bird cluster.\\
\begin{lstlisting}[caption= Synchronization APIs, label=code:api2]
class DBarrier : public DObject{
	DBarrier(int count);
	bool Enter(int timeout=-1);
};
class DSemaphore : public DObject{
	DSemaphore(int count);
	bool Acquire(int timeout=-1);
	void Release();
};
\end{lstlisting}

The \texttt{DBarrier} class provides barrier synchronization pattern to keep distributed threads (i.e., main thread and working threads) in the same pace, which is useful in performing synchronous iterative computation. Typically, a \texttt{DBarrier} object is created by the \emph{main thread} in \bird master. The constructor in \texttt{DBarrier} is then invoked to create a barrier and specify the total number of threads to be synchronized on the barrier. The reference to a \texttt{DBarrier} object can be stored in a shared global variable so that all threads in the cluster can share this barrier. After setting up all the working threads in slaves, the \emph{main thread} calls \texttt{Enter} function and waits at the barrier until all the working threads reach the barrier. When the last thread arrives at the barrier, all the threads will resume their normal execution. 

The \texttt{DSemaphore} class allows a specified number of threads to access a resource. During the creation of a \texttt{RSemaphore} object, we set a non-negative \emph{resource count} as its value. There are two ways to manipulate a semaphore. \texttt{Acquire} function is used to request a resource and auto-decrement the \emph{resource count}; \texttt{Release} function is used to release the resource and auto-increment the \emph{resource count}. Threads that request a resource with non-positive semaphore value will be blocked until other threads release that resource and the semaphore value becomes positive. 

The above synchronization interfaces provide basic building blocks for user applications and are designed to be \emph{aligned} to Pthreads APIs.
With this similarity, developers who are familiar with traditional multi-threaded programing are able to perform distributed multi-threading with \bird effectively.

\subsection{Accumulator}\label{sec:api:accu}
We found that many real applications require to perform vector-wise accumulation.
For example, in PageRank computation, each working thread maintains a subset of vertices with their outgoing edges. During each iteration, a working thread computes the PageRank credits from its own vertices to the destination vertices along the edges. The credit vectors from different working threads are summed together to produce new PageRank values for all vertices.

A straightforward way to perform vector accumulation with \bird is to ask working threads to transfer local vectors to DSM, and then choose one thread to fetch all vectors, perform final accumulation locally, and forward newly computed vector elements to DSM or the corresponding threads. 
Let $N$ be the number of working threads. The above method incurs high network cost, i.e., the size of data to be transferred is at least $(2N+1)*vector\_size$.

\bird provides \texttt{DAddAccumulator} class for users to perform vector accumulation more efficiently as well as hide data transfer details involved in vector accumulation.
Users can create a shared \texttt{DAddAccumulator} object and initialize it with the number $N$ of working threads and an output shared array in DSM.
The working threads can invoke \texttt{Accumulate} function defined in \texttt{DAddAccumulator} to send out their local vectors and \bird will compute the final accumulated result and store it in the output shared array automatically. {The \texttt{Accumulate} function also acts like a synchronization point which will not return until all the $N$ threads send out their local vectors.}
Our implementation of \texttt{Accumulate} function reduces the data transfer cost to $(N+1)*vector\_size$ (see details in Section~\ref{impl:accumulator}).

\subsection{Example: Putting {Them} All Together}\label{subsec:prog:lr}

Now we illustrate how to develop applications with \bird interfaces using logistic regression as an example.
Logistic regression~\cite{prml} is a widely used discriminative model for two-class classification\footnote{We consider binary logistic regression for simplicity. The solution can be easily extended to multinomial case.}.
Given a $d$-dimensional explanatory vector {\bf x}, the logistic regression model is to predict the probability of a binary response variable $y$ taking 0 or 1, based on logistic function.
That is, $p(y=1|{\bf x})= \sigma(\theta^{T}{\bf x}) =\frac{1}{1+\exp(-\theta^{T}{\bf x})}$,
where $\theta\in R^d$ is a parameter vector and $\sigma$ is the logistic function.
The objective of logistic regression is to learn $\theta$ that maximizes the conditional log likelihood of training data $\{({\bf x}_i, y_i)| i\in[1,n]\}$.
\eat{That is, $p(y=1|{\bf x})= \sigma(\theta^{T}{\bf x}) =\frac{1}{1+\exp(-\theta^{T}{\bf x})}$,
	where $\theta\in R^d$ is a parameter vector and $\sigma$ is the logistic function.
	The objective of logistic regression is to learn $\theta$ that maximizes the conditional log likelihood of the training data $\{({\bf x}_i, y_i)| i\in[1,n]\}$: $\max_{\theta} \sum_{i=1}^n {\log p(y_i|{\bf x}_i, \theta)}$.
	
	We adopt the mini-batch stochastic gradient descent (SGD) algorithm to learn $\theta$, which is effective in handling sheer amount of data~\cite{sgd}.
	Specifically, we update $\theta$ iteratively using the following update function:
	\begin{equation}\label{eq:grad}
	\small
	\delta =  \alpha \sum_{j\in B}{ \big(y_j - \sigma(\theta^T {\bf x}_j)\big){\bf x}_j }
	\end{equation}
	\begin{equation}\label{eq:update}
	\small
	\theta^{\bf new} = \theta + \delta
	\end{equation}
	where $\alpha$ is the learning rate and $B\subseteq \{1,\cdots,n\}$ is a mini-batch of training data. 
}
We adopt the mini-batch stochastic gradient descent (SGD) algorithm~\cite{sgd} that updates $\theta$ iteratively using the following update function.
\begin{equation}\label{eq:update}
\small
\theta^{(i+1)} = \theta^{(i)} + \alpha \delta
\end{equation}
where $\alpha$ is the step size, and $\delta=\sum_{p\in B}{ \big(y_p - \sigma(\theta^{(i)T} {\bf x}_p)\big){\bf x}_p }$ is the gradient over mini-batch $B$ of training data.

The \bird program for the entry function (i.e., {\sf thread\_proc} function) of working threads and the shared data declaration in logistic regression is provided as follows. We omit initialization and finalization details for simplicity. 
\\
\begin{lstlisting}[numbers=left,stepnumber=1,xleftmargin=2.5em,frame=single,framexleftmargin=2.5em]
struct DataPoint{
	float y;
	float *x;
};
DefGlobal(param_len, int);
DefGlobal(grad, Array<float>);
DefGlobal(accu, Ref<DAddAccumulator<float>>);

void slave_proc(uint32_t tid){
	float* theta = InitialParam();
	std::vector<DataPoint> points =LoadTrainPoint(tid);
	float* local_grad = new float[param_len];
	for (int i = 0; i < ITERATIONS; i++){		
		std::fill_n(local_grad, param_len, 0);
		for (auto p : points){
		float dot = 0;
		for (int j = 0; j < param_len; j++)
			dot += theta[j] * p.x[j];
			for (int j = 0; j < param_len; j++)
			local_grad[j] += (p.y -  1 / (1 + exp(-dot))) * p.y;	
		}
		accu->Accumulate(local_grad, param_len);
		for (int j = 0; j < param_len; j++)
			theta[j] += step_size * grad[j];
	}
	//finalization code
	...
}
\end{lstlisting}

In our design, every working thread maintains the parameter vector $\theta$ in a local array {\sf theta} (line $10$) whose length is stored in a shared variable {\sf param\_len}. 
We use a shared array {\sf grad} in DSM to keep the global gradient vector, and associate {\sf grad} with an accumulator {\sf accu} that sums over thread-level gradients {\sf local\_grad} and stores the accumulated result to {\sf grad}.
We partition the training set into disjoint mini-batches and assign them to the working threads uniformly via a user-defined partition function \texttt{LoadTrainPoint} (line $11$). 
In each iteration, every working thread computes local gradient {\sf local\_grad} based on its mini-batch training data (line $15$-$21$), and the accumulator {\sf accu} is used to sum up all local gradients {\sf local\_grad} (line $22$) to compute the global gradient {\sf grad}. Finally, each working thread updates its local parameter vector {\sf theta} by adding {\sf step\_size}*{\sf grad} with {\sf theta} (line $23$-$24$). Such computation is repeated by {\sf ITERATION} times.


\eat{\stitle{Comparison with Spark.} Spark's paper~\cite{spark} provides an example for logistic regression, which is shown below. 
	\begin{lstlisting}[basicstyle=\ttfamily\scriptsize,
	language=Scala,
	tabsize=2,
	breaklines=true,
	frame=single,
	aboveskip=0em,
	belowskip=0em,
	captionpos=b]
	// Read points from a text file and cache them
	val points = spark.textFile(...)
	.map(parsePoint).cache()
	// Initialize w to random D-dimensional vector
	var theta = Vector.random(D)
	// Run multiple iterations to update theta
	for (i <- 1 to ITERATIONS) {
	val grad = spark.accumulator(new Vector(D))
	for (p <- points) { // Runs in parallel
	val s = (1/(1+exp(-p.y*(theta dot p.x)))-1)*p.y
	grad += s * p.x
	}
	theta -= grad.value
	}
	\end{lstlisting}
	The Spark's example has shorter code length than \bird's. We attribute this to two reasons. Firstly, Spark and its underlying language Scala has ``syntactic sugar'' for vector operations, including operators to calculate the sum and dot product for vectors. This extensively minimized the code length. Secondly, Spark adopts implicit parallelism, where developers do not need to explicitly create, manage, and synchronize threads. Implicit parallelism brings convenience for developers. However, we show in the later part of this section that we can optimize the code by explicitly managing threads.
}

{\bf Discussion.} 
Various implementations of {\sf thread\_proc} function (i.e., {\sf slave\_proc} in the above example) can be adopted to further optimize the performance for logistic regression. 
For instance, we find that the global gradient \texttt{grad} is fetched by all threads in the same node (line $24$). One can improve the example code by letting only one thread in each slave fetch the global gradient and share it with other threads via a local array (since threads within the same node share the local memory space).
Moreover, one can use a single thread in each slave to combine all local gradients from that node and then accumulate the combined results via the accumulator.
Both optimizations can help reduce data transfer cost between DSM and local memory.
Note that the above fine-grained optimizations can hardly be achieved using programming primitives from existing general-purpose distributed systems; and more importantly, the \bird interfaces are useful to achieve these optimizations in a natural and efficient way.

\section{Implementation Details} \label{impl}
\eat{
	\begin{figure}[t]
		\centering
		\includegraphics[width=0.8\linewidth]{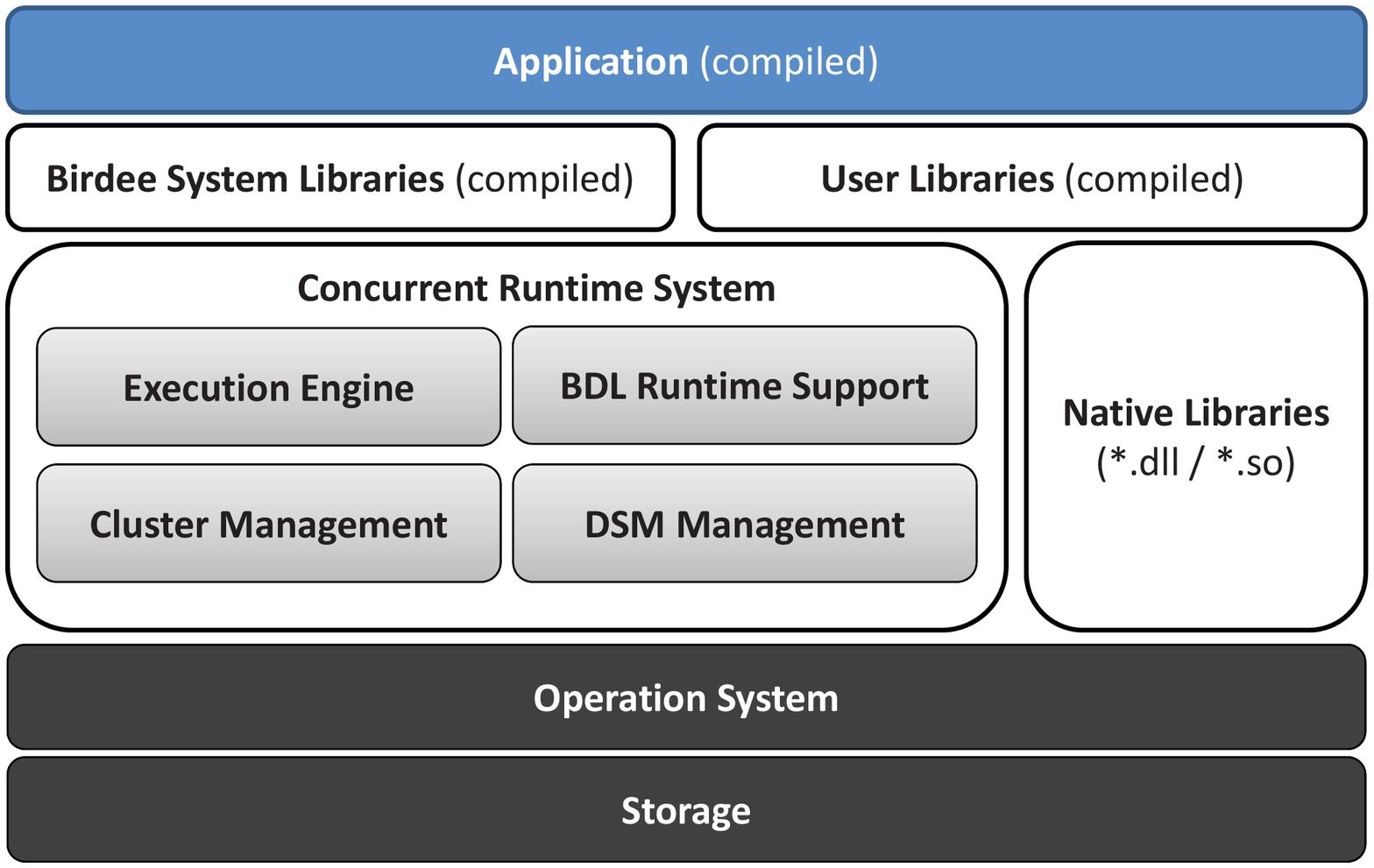}
		\caption{\small Runtime stack of \bl}
		\label{fig:stack}
		\vspace{-.1in}
	\end{figure}
}

\eat{
	Since \bl and \blib are two developer interfaces for \bird, they share the same underlying the distributed system library. Furthermore, we implement a compiler and runtime support for \bl only, while \blib is supported by the existing C++ compiler and C++ runtime.
}
The following subsections discuss the technical details of \bird implementation\footnote{\small{The full code of \bird is available in \url{https://github.com/STEP-dev/STEP}}}.

\subsection{Distributed Shared Memory Management} \label{subsec:dsm}

\begin{figure}[t]
	\centering
	\includegraphics[width=.62\linewidth,height=3.7cm]{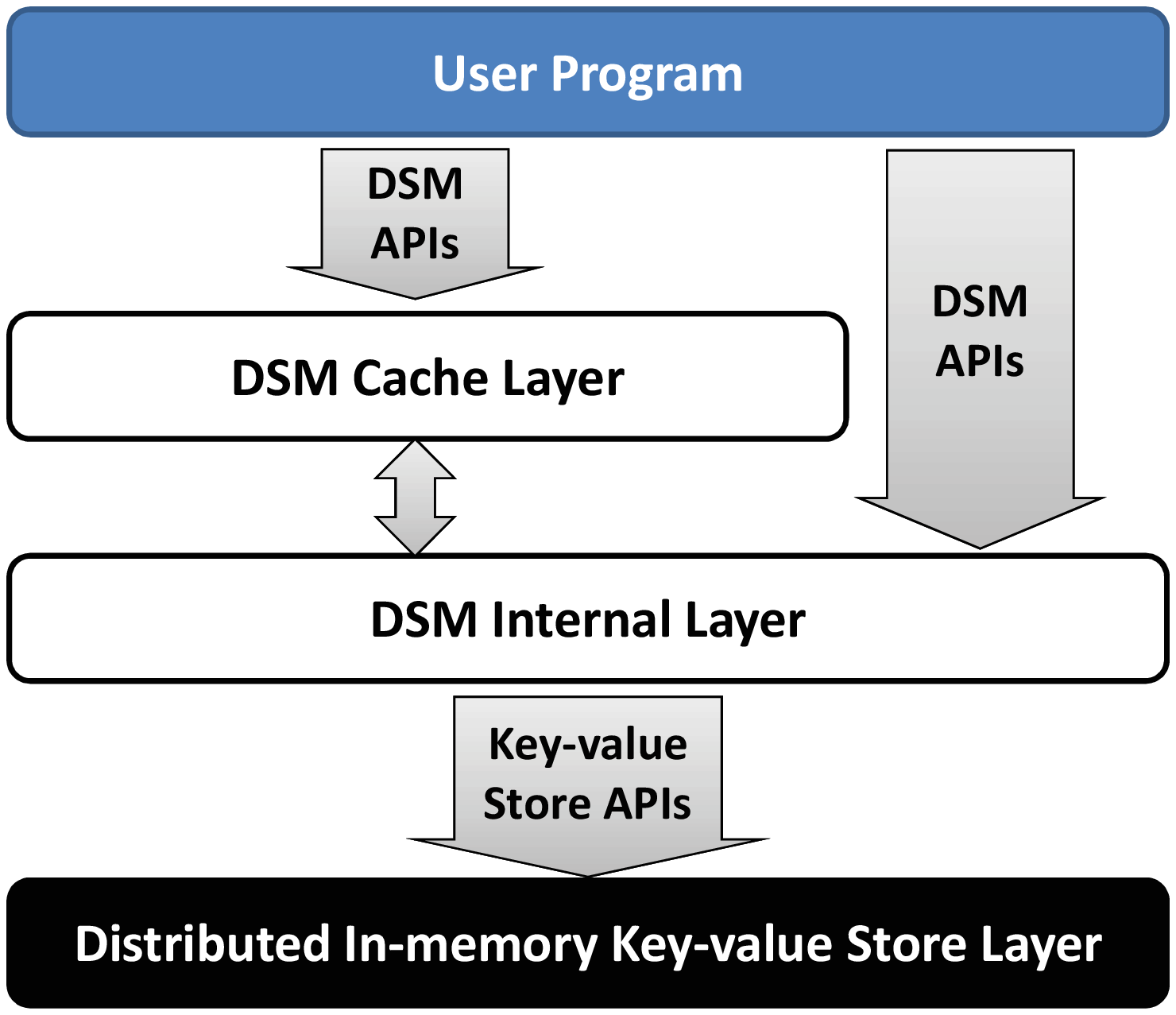}
	\vspace{-.1in}
	\caption{\small DSM management in \bird}
	\label{fig:dsm}
	\vspace{-.2in}
\end{figure}

\bird provides distributed shared memory (DSM) to store globally shared data among threads in \bird cluster. We implement DSM following a three-layer architecture, as shown in Figure~\ref{fig:dsm}. 

$\bullet$ The bottom layer contains the off-the-shelf distributed in-memory key-value store which keeps all shared data physically. We use memcached~\cite{memcached}, a simple yet powerful object caching system, in our current implementation. 
We associate each piece of shared data with a unique key and store the (key, shared data) pair into memcached. All the requests sent to the bottom layer are key-value store specific. 

$\bullet$ The middle layer, named DSM internal layer, separates user-level DSM data access from specific key-value store implementation. Particularly, it handles unified DSM API calls (e.g. \texttt{Get()} and  \texttt{Set()}) from \bird programs, and transforms them into the operations provided by the underlying key-value store. This transformation is important in face of the rapid evolution of key-value stores. That is, we can easily switch to another more efficient key-value store without any modification in user programs. 

$\bullet$ The top layer is a DSM cache that leverages spatial and temporal locality to facilitate fast shared data access in DSM. 
We implement a directory-based distributed memory cache in this layer. 
It absorbs DSM data access when there is a cache hit. We also allow some DSM API calls to skip DSM cache layer atomic operations on the shared data, e.g., atomic-increment or atomic-decrement on shared counters. 

In what follows, we first present our design of shared memory address space in \bird and then provide implementation details in DSM management.

\eat{
	Birdee's shared memory space resides in DSM, making it an important part of Birdee's distributed system. We adopt a four-level abstraction in the design of DSM. The top level exposes all the interfaces a Birdee user program needs, including interfaces for reading from and writing to DSM, as well as incrementing and decrementing an atomic counter in DSM. The second level is the DSM cache level. We have implemented a directory-based distributed memory cache here. The third level, which is the abstraction level, provides a stable set of interfaces for the upper levels and hides the implementation of the lowest level of the DSM. For the time being, we choose the distributed in-memory key-value database, memcached, to implement the lowest level of Birdee's DSM. Due to the abstraction level of DSM, the change of the implementation of the lowest level does not affect the higher levels. Therefore, we can switch to a more efficient DSM module if possible without a need to modify the code of other levels. Note that some of the interfaces of the top-level, including operations on atomic counters, directly call the corresponding interfaces of the abstraction level, skipping the cache level.\\
	This section will then discuss some important designs of Birdee's DSM.
}

{\bf Shared Memory Address Space.}
\bird allows 64-bit shared memory addresses and organizes the complete DSM space in the granularity of \textbf{word}, i.e., 32 bits. 
That is, every 64-bit shared memory address identifies a 32-bit chunk in DSM. \bird interprets each shared memory address as a high-order $x$-bit {\sf object\_id} plus a low-order $(64-x)$-bit {\sf field\_id}. A 32-bit word and 64-bit address size can support up to $2^{64}*4$ bytes DSM space. {For a data type with over 32 bits, any of its instances will occupy multiple words in DSM.}
By default, we set $x$ to $32$, which allows \bird to support up to $2^{32}$ shared {\sf object}s and $2^{32}$ {\sf field}s for each of them.

To access a field of an object in DSM, \bird runtime system will fetch its {\sf object\_id} and {\sf field\_id} to compose a 64-bit shared memory address. This address will be used as a \emph{key} to access the object's field value in key-value store. 
In addition to objects, \bird also allows arrays and variables to be stored in DSM. The {\sf object\_id} for an array is a unique array identifier and the {\sf field\_id} {is the index of an array element. }
For shared variables, \bird runtime allocates a unique {\sf field\_id} for each of them. \bird system has a virtual object with {\sf object\_id} equal to 0, which holds all shared variables in the program.

The above interpretation of shared memory addresses allows fields in the same object (or elements in the same array, or global variables) to be stored in continuous shared memory space, which enables DSM cache layer to exploit spatial locality and reduce shared data access latency.

\eat{
	A \textbf{word} in the shared memory space of Birdee is a 64-bit long chunk of data, which is the granularity of Birdee's DSM. One can locate and access a 64-bit-long word using a 64-bit long shared memory space address. A word is large enough to store a 32-bit integer, a double-precision float-point number or a shared memory space address.\\
	The shared memory space address is split into two parts, which are the higher 32 bits and the lower 32 bits. The higher 32 bits are called object-id, which is used to identify which object the word belongs to. The lower 32 bits, on the other hand, is called field-id, identifying which field of the object the word is. To access a field of an object in DSM, the Birdee runtime first fetches the object-id and the field-id of the field. Then it appends the 32-bit field-id to 32-bit the object-id, making a 64-bit shared memory space address. Finally, by using the address, we have the access to the field. Thanks to the addressing design, we have a good special locality for caches because the addresses of all fields of the same object have the identical higher 32 bits. Our design of the address space makes our framework support creating at most 2\textsuperscript{32} objects or arrays and every object or array has at most 2\textsuperscript{32} fields or elements.
}

{\bf Key-valure Store Layer.}\label{sec:kv}
\bird implements key-value store layer based on libmemcached~\cite{memcachelib}, an open-source C client library and tool for memcached server.
Note that memcached is originally designed as a fully-associative distributed cache and will discard the oldest data periodically or when running out-of-memory. 
To address the problem, we disable automatic cache data eviction in memcached via appropriate system parameter settings. However, when the cache is full, subsequent insertions will still throw out-of-memory exceptions. This limitation can be solved by leveraging persistence enabled key-value stores such as memcacheDB~\cite{memcacheDB}, which we leave as future work.

\bird stores all shared data as key-value pairs in memcached where every {\bf key} is a 64-bit shared memory address and the {\bf value} contains a word-sized chunk. 
We refer to this implementation as {\bf fine-grained DSM}. However, fine-grained DSM achieves merely 33\% effective key-value store usage due to the fact that only the value part in a pair is used to store real shared data and the key part is typically unused.
Furthermore, in fine-grained DSM, reads and writes to shared data in large sizes may involve a number of network requests and thus is inefficient in terms of data transmission between local memory and DSM. 

In face of the above problem, \bird introduces {\bf coarse-grained DSM} that associates a shared memory address (i.e., the key) with several consecutive words, called a \emph{package}, and stores (key, package) as a key-value pair in memcached.
By default, a package contains 32 words and \bird guarantees that key-value pairs are stored at package-size-aligned shared memory addresses.  
Coarse-grained DSM reduces the number of data transmission requests for large-size shared data access and hence improves the overall DSM throughput. Unfortunately, this solution will increase DSM access latency. That is, an update to a single word in the package requires to access the whole package from DSM.

\bird allows developers to decide which DSM mode to use via \bird configuration file. 
According to our experimental results, coarse-grained DSM achieves better performance than fine-grained DSM in many real applications.

\eat{
	\begin{table}
		\centering
		\caption{\small DSM APIs}\label{tab:dsm}
		\scriptsize
		\begin{tabular}{|c|c|c|}  
			\hline
			\bf Types & \bf Read from DSM & \bf Write to DSM \\\hline
			Integer & \texttt{SGeti()} & \texttt{SSeti()}\\ \hline
			Double & \texttt{SGetd()} & \texttt{SSetd()}\\	\hline
			Float & \texttt{SGetf()} & \texttt{SSetf()}\\\hline
			Reference & \texttt{SGeto()} & \texttt{SSeto()}\\\hline
		\end{tabular}
	\end{table}
}
{\bf DSM Internal Layer.}
DSM internal layer provides a set of DSM APIs functions for getting and setting values of shared data of user programs on DSM.
Specifically, users access \bird shared data in the same way as operating data in local memory (recall the details in Section~\ref{sec:bl}) and
\bird library transforms all DSM data accesses into DSM API calls. 
We want to emphasize that DSM APIs are invisible to users and only accessible to \bird library internally.

\bird implements all DSM APIs based on the interfaces from the underlying key-value store (see Table~\ref{tab:kv}).  
Typically, DSM APIs include two kinds of operations: setting or getting shared data given its address. 
All the \emph{set} functions have three parameters. The first and second parameters represent the {\sf object\_id} and {\sf field\_id} of the shared data, respectively. Concatenating these two parameters produces a 64-bit shared memory address to locate the data.
The third parameter is the updated data value. 
The address-value pair will be passed to the underlying key-value store API functions, e.g., \texttt{memcached\_set()} in memcached.  

Similarly, all the \emph{get} functions involve two parameters representing the {\sf object\_id} and {\sf field\_id} of the shared data, respectively. In \emph{get} functions, we first concatenate the input parameters to compose a shared memory address. We then call the corresponding key-value store API method (e.g., \texttt{memcached\_get()} in memcached) with the shared memory address as the \emph{key}. the Finally, the \emph{get} functions return the data fetched by the key-value store.
%


{\bf DSM Cache Layer.}
The development of DSM cache layer is motivated by two important observations. First, frequent shared data access incurs a large amount of network communication cost. Second, the performance of \bird is compromised by the limited throughput of the underlying key-value store. 
Hence, we implement a write-through distributed DSM cache with the purpose of reducing networking cost and alleviating the burden of key-value store.

We organize both DSM and DSM cache into \emph{blocks} where each block contains 32 words. 
DSM has $2^{59}$ \emph{data blocks} using 64-bit shared memory addresses. The high-order 59 bits of a shared memory address represent the \textbf{address} of its belonging data block. 
Every node in a \bird cluster is designed to contain 1024 DSM \emph{cache blocks}. 
DSM cache adopts LRU strategy for block replacement. That is, when all cache blocks in the same node are used, we evict the block that is unused for the longest time.

\bird guarantees DSM cache coherence with the directory-based protocol~\cite{Hennessy:2003:CAQ:861856}.
Let $n$ be the total number of nodes in \bird cluster.  We require a node to be the \textbf{watcher node} for a data block iff 
$$\small node~id \equiv block~address~~(\bmod~n).$$
Note that each data block is watched by exactly one node. In \bird, every watcher node maintains a directory recording which nodes have a copy of its watching data blocks.

When a node calls DSM API to read data in a shared memory address, \bird runtime first searches local cache blocks. If the cache hits, the thread is able to retrieve data directly without introducing network cost. Otherwise, \bird runtime forwards the DSM API call to the DSM internal layer and sends a ``missing'' message to the corresponding watcher node for the required data block. The watcher node receives the message and updates directory for the data block accordingly.

For a DSM write, \bird adopts write-invalidate protocol. \bird runtime will first check all the local cache blocks. If there is a cache hit, the writing thread can perform local update and send a ``write'' message to the watcher node for the updated data block.
Once receiving the ``write'' message, the watcher node refers to the directory and sends an ``update'' message to all the nodes that cache the block. The ``update'' message will be used to invalidate the stale copies of data blocks in DSM cache.

\subsection{Accumulator}\label{impl:accumulator}
The accumulator is designed to reduce the network traffic incurred by accumulating multiple vectors.
Let $N$ be the number of working threads that hold local vectors to be accumulated and $M$ be the number of slaves in the cluster. When a thread invokes \texttt{Accumulate} function defined in \texttt{DAddAccumulator},
it divides the local vector into $M$ chunks and forwards the $i$-th chunk to the node with node ID of $i$.
Upon receiving all chunks, a node performs accumulation over the sub-vectors in local memory and updates the corresponding elements in the output array. This method reduces the total amount of transferred data to $(N+1)*vector\_size$.

In some applications, the vectors to be accumulated are \emph{sparse}, 
i.e., with few non-zero elements. Hence, \bird provides three modes of accumulator: \emph{dense, sparse, auto}. The \emph{dense} mode behaves as described above. 
In \emph{sparse} mode, a vector is represented by (index, non-zero element) pairs.
Transferring pairs for non-zero elements incurs lower network cost if the vector is sparse. 
In \emph{auto} mode, \bird checks if it is beneficial to convert vectors to the pairs and automatically chooses the mode with lower network cost.

\subsection{Synchronization}

We implemented two thread synchronization mechanisms in \bird. 
The first mechanism is \emph{barrier}. Specifically, \bird master maintains a counter 
for each distributed barrier in \bird cluster. When a thread (main or working) enters the barrier, it sends a message to the \emph{sync controller} on master node to increase the counter by 1. 
Every thread entering the barrier waits for the release of the barrier. 
When the counter reaches the threshold defined on barrier creation, \emph{sync controller} broadcasts a ``release'' command to all the threads blocked by this barrier. The \emph{synchronizer} of each slave node then resumes the execution of the blocked threads.

The second synchronization mechanism is \emph{semaphore}.
The implementation of semaphore is similar to that of barrier. 
The \emph{sync controller} in \bird master manages a counter upon the creation of a distributed semaphore. 
When a thread \texttt{Acquire}s control over a semaphore, \emph{sync controller} will check the count of the semaphore. If the count is non-positive, it puts the thread into a waiting queue for the semaphore. Otherwise, the counter is decreased by 1 and the thread is left unblocked. When a thread holding the semaphore calls \texttt{Release} method,  \emph{sync controller} will increase the counter by 1. 
When the counter becomes positive, \emph{sync controller} resumes the execution of the first thread in the waiting queue (in FIFO manner) if any, and auto-decrements the counter.

\subsection{Fault Tolerance}


\bird leverages heartbeat messages to detect node failures.
That is, every slave node is requested to send heartbeats to \bird master periodically. 
If the master does not receive the heartbeat message from a slave over a fixed time interval, the slave is considered to be dead. Upon failures, \bird master will send ``recovery'' message to all the remaining slaves and the recovery process starts immediately. 

\bird adopts checkpoint-based recovery mechanism.
For synchronous iterative applications, we make checkpoints every a few iterations. 
This is achieved by inserting appropriate \emph{checkpoint logic} right before the barrier is released. 
When doing a checkpoint, \bird uses fault-tolerant distributed file system to maintain a consistent copy of the data in DSM and any important information to be materialized.
During recovery, \bird master creates new working threads in healthy nodes to replace the failed ones.
All the threads then rollback to the latest checkpoint, re-load input data if necessary and redo the computation with the latest consistent copy of DSM.
Besides, \bird provides an abstract \texttt{Checkpoint} class with two functions \texttt{DoCheckpoint()} and \texttt{DoRestart()} for users to store extra information for recovery. Users may extend \texttt{Checkpoint} class to specify the variables to be persisted during checkpointing. They can also transform program-specific state to a \texttt{Checkpoint} instance and vice versa, while 
\bird is responsible for invoking \texttt{DoCheckpoint()} during checkpointing (to persist \texttt{Checkpoint} instance)
and calling \texttt{DoRestart()} during recovery (to restore program state from the materialized \texttt{Checkpoint} instance).
Doing checkpointing for non-iterative (or asynchronous) computation tasks is more subtle because no barrier is available. 
To address the problem, \bird master is able to send a ``checkpoint'' command to pause the execution of all slaves. This command is used to enforce a virtual barrier. Upon receiving the command, every slave stops all its working threads (after finishing the computation in hand) to create a checkpoint. Similarly, users can leverage \texttt{Checkpoint} class to indicate a consistent state of the program that will be persisted by \bird automatically.

\section{Experimental Evaluation} \label{sec:exp}

%
%
%

\subsection{Experiment Setup}\label{sec:setup}
\eat{the cluster based on four physical machines. 
	Each of them has a 64-core Intel Xeon E5-4610 CPU v2 clocked at 2.30GHz, 64GB of memory and 16TB SATA hard disks. All the physical machines run CentOS 6.6 operating system and are connected via a 1 Gbps network switch.
	We set up four virtual machines on every physical machine and obtain}
\eat{. We use kernel-based virtual machines\footnote{\small\url{http://www.linux-kvm.org/page/Main_Page}} (KVM) to construct a 16-node cluster.Each node is equipped with a four-core virtual CPU, 4GB of memory and 2TB hard disk, running Ubuntu 14.04 x64.}

The experimental study was conducted on our in-house cluster. The cluster consists of 16 Dell M630 compute nodes, each of which is equipped with one Intel E5-2609-V3 CPU, 64GB RAM and 600GB hard disk, running CentOS 6.6 operating system. All the nodes are hosted on one rack and connected by a 10Gbps switch.

We compare \bird with two Spark extensions -- MLlib~\cite{mllib} and GraphX~\cite{graphx}, Husky~\cite{husky} and a specialized machine learning platform Petuum~\cite{petuum}.
We installed version-2.1.1 of Spark and the latest versions of Petuum and Husky. We also carefully tuned the settings to get the best system performance. 
For \bird, we use auto mode for the accumulator and coarse-grained DSM mode by default.

\subsection{Applications and Datasets}
\begin{table}[t]
	\vspace{-.1in}
	\centering
	\small
	\caption{\small Code lengths in available implementations }\label{tab:impl}
	\begin{tabular}{|c|c|c|c|c|}\hline
		& \bird & Petuum & Husky & Spark \\ \hline
		Logistic Regression & 323 & - & -  & 213\\ \hline
		K-means & 285 & 1446 & - & 372 \\ \hline
		NMF & 311 & 1144 & - & - \\ \hline		
		PageRank & 279 & - & 107 & 215 \\ \hline	
	\end{tabular}
	\vspace{-.1in}
\end{table}

{\bf Applications.}
We evaluate the performance of different distributed systems using four applications: logistic regression, K-means, non-negative matrix factorization (NMF) and PageRank.
These applications are considered to capture different workloads: machine learning tasks, graph analytics tasks, requiring little or substantial amount of shared data to be managed.
We implemented all the applications in \bird system. For Spark, Petuum and Husky, we directly used the implementations from the libraries or examples shipped with these open-source systems.
We guarantee that each application is implemented using the same algorithm in different systems.
For instance, both \bird and Spark adopt the mini-batch SGD algorithm for logistic regression.

Table~\ref{tab:impl} shows the available implementations for the applications and the corresponding code lengths provided by MLlib\footnote{\url{https://spark.apache.org/docs/2.1.1/mllib-guide.html}} and GraphX\footnote{\url{https://spark.apache.org/docs/2.1.1/graphx-programming-guide.html}} in Spark, Petuum Bosen\footnote{\url{https://github.com/petuum/bosen/tree/stable/app}} and Husky\footnote{\url{https://github.com/husky-team/husky}}.
For \bird applications, each of them is written in one source file using C++.
Spark applications are implemented in the corresponding RDDs. 
%
We measure the code lengths of {\sf KMeans.scala}, {\sf LogisticRegression.scala} and {\sf PageRank.scala} from Spark packages excluding the comments and unrelated code.
For Petuum, we consider the codes in {\sf .cpp, .hpp, .h} files for each application under the {\sf app} directory and exclude the relevant codes from Petuum's native library. 
We count the lines of {\sf .cpp} files for Husky PageRank application under the directory of {\sf benchmarks} with relevant system library source codes excluded.

We can see that \bird applications require comparable code lengths compared with Spark and Petuum over all the applications, which illustrates the effectiveness of programming with \bird. 
Husky requires the shortest code length for PageRank. This is because Husky utilizes {\sf Boost} library to parse input data, leading to shorter code length in Husky.

{\bf Datasets.}
Table~\ref{tab:dataset} provides a detailed description for all the datasets used throughout the experiments. 

$\bullet$ \textbf{Logistic regression.} 
We consider two datasets for logistic regression: GENE and LRS.
GENE~\cite{parkinson2009arrayexpress} is a gene-expression dataset (accession number: E-TABM-185). It contains 22K data rows over about 6K features representing different cell lines, tissue types, etc. 
LRS\footnote{\small\url{http://finzi.psych.upenn.edu/library/COUNT/html/logit_syn.html}} is a synthetic dataset generated by the COUNT library in R language. LRS contains 30K features and we use LRS to evaluate the effects of feature dimensionality on system performance.

$\bullet$ \textbf{K-means}. 
We use two datasets for K-means computation: FOREST and KMS.
FOREST\footnote{\small\url{http://archive.ics.uci.edu/ml/datasets/Covertype}} contains forest cover types for observations (30x30 meter cell) from US Forest Service Region 2 Resource Information System. Each observation contains 54 qualitative independent variables (e.g., soil types) as features.
KMS\footnote{\small\url{https://github.com/petuum/bosen/blob/stable/app/kmeans/dataset/data_generate.py}} is a synthetic dataset generated by the generator script from Petuum's K-means package.
We set the number of features to 4096 and obtain 240K data points using the script.

$\bullet$ \textbf{NMF}.
We use two datasets for NMF. NETFLIX~\cite{bennett2007netflix} is a movie rating dataset with 38K ratings. Each rating is associated with 17K features. NMFS is a synthetic matrix where [i,j]-th element equals i$\times$8K+j. The matrix contains 8K features (i.e., columns) and 512K data rows.

$\bullet$ \textbf{PageRank}.
We conduct PageRank computation using two real datasets. LJ and FRIEND, both of which are online social networks. 
LJ\footnote{\small\url{https://snap.stanford.edu/data/com-LiveJournal.html}} contains over 4 million vertices and 70 million directed edges. 
FRIEND\footnote{\small\url{https://snap.stanford.edu/data/com-Friendster.html}} includes more than 60 million vertices and 1 billion directed edges.

\begin{table}[t]
	\vspace{-.05in}
	\centering
	\caption{\small Dataset descriptions}\label{tab:dataset}
	\scriptsize
	\begin{tabular}{|l|l|l|l|}  \hline			
		& \bf Datasets & \bf \#features & \bf \#data rows\\\hline
		\multirow{2}{*}{Logistic regression} & GENE & 5896 & 22283\\	
		& LRS & 30720 & 30000\\	\hline 
		\multirow{2}{*}{K-means} & FOREST & 54 & 581012\\ 
		& KMS & 4096 & 200000\\	\hline
		\multirow{2}{*}{NMF} & NETFLIX & 17700 & 384000\\ 
		& NMFS & 8192 & 500000\\\hline
		\hline
		& \bf Datasets & \bf \#vertices & \bf \#edges\\\hline
		\multirow{2}{*}{PageRank} & LJ & 3997962 & 34681189 \\	
		& FRIEND &65608366& 1806067135  \\	\hline
	\end{tabular}
	\vspace{-.2in}
\end{table}

{\bf Parameter settings and metric.}
We try different values for the number of clusters in K-means, the factorization rank for NMF, the number of iterations performed and the number of nodes in the cluster.
Table~\ref{tab:param} summarizes the ranges of our tuning parameters. 
Unless otherwise
specified, we use the underlined default values. The staleness of tables (SSP) in Petuum is set to zero by default.
We measure the running time for the systems over different applications. All the results are averaged over ten runs. 

\begin{table}[t]
	\centering
	\scriptsize
	\caption{\small Parameter ranges}\label{tab:param}
	\begin{tabular}{|l|l|l|}
		\hline
		& \bf Parameter & \bf Range \\\hline
		\multirow{2}{*}{K-means} & \#clusters(FOREST) & $ 10,20,\underline{30},40,50 $  \\
		& \#clusters(KMS) & $ 16, 32, 64 , 128 ,\underline{256}$ \\
		\hline
		NMF & Factorization rank & $ 8,16,\underline{32},64,128$ \\
		\hline
		ALL\footnote{\scriptsize We try 10, 20, 30, 40, 50 iterations for K-means on FOREST due to its short running time per iteration} & \#iterations &$6, 8, \underline{10}, 12, 14$ \\
		\hline			
	\end{tabular}
	\vspace{-.1in}
\end{table}

\eat{
	\subsection{Dense Versus Sparse Accumulator}
	
	\begin{figure}[t]
		\centering
		\subfigure[\small LJ dataset]{
			\label{fig:pr_lj:accu}
			\epsfig{height=2.9cm,width=0.49\linewidth,file=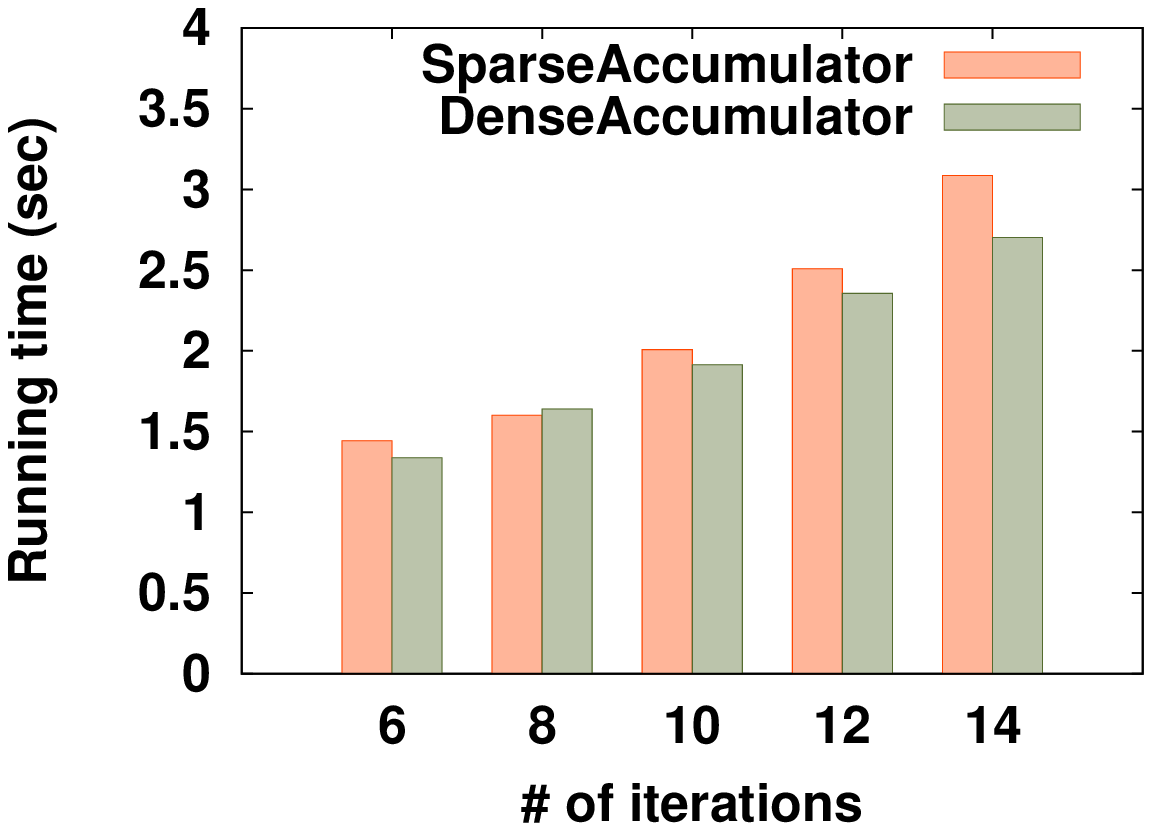}}
		\subfigure[\small FRIEND dataset]{
			\label{fig:pr_friend:accu}
			\hspace{-2mm}\epsfig{height=2.9cm,width=0.49\linewidth,file=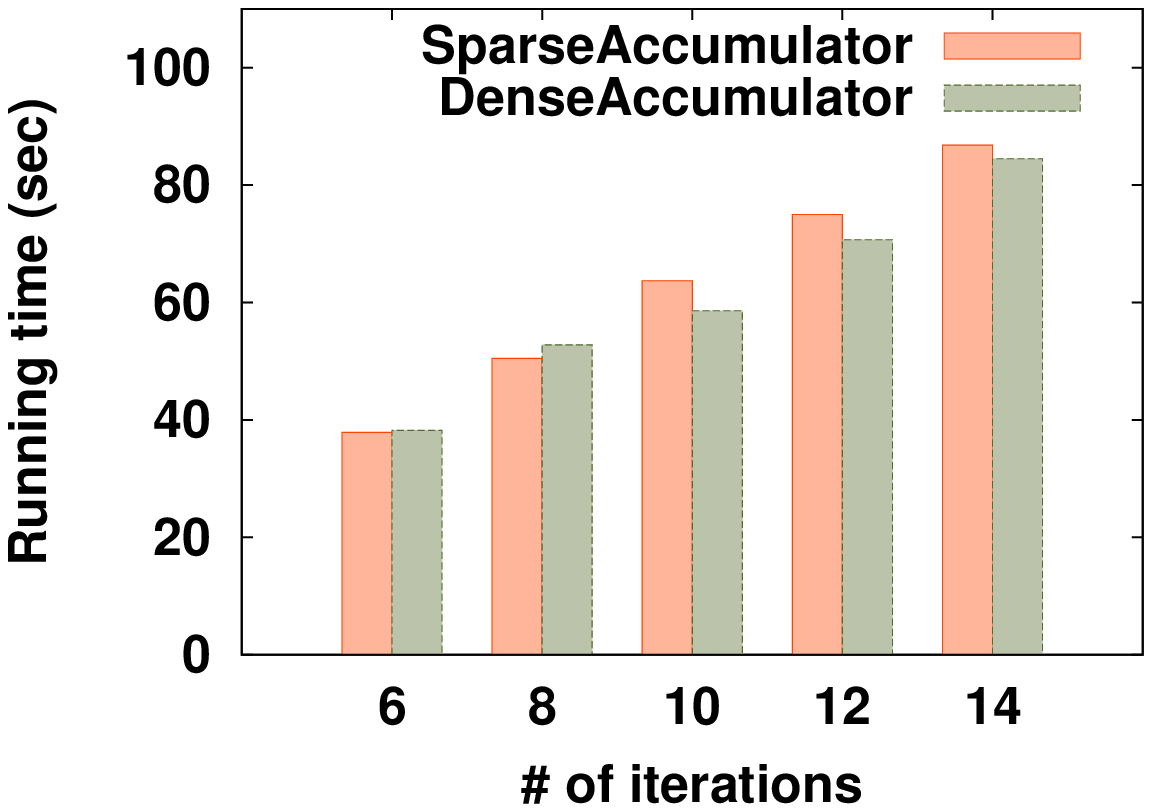}}
		\vspace{-.1in}
		\caption{\small Dense vs sparse accumulator (PageRank)}
		\label{fig:pr_accu}
	\end{figure}
	Figure~\ref{fig:pr_accu} shows the comparison result of different accumulators using PageRank.
	\bird PageRank implementation with {\sc SparseAccum} runs $1.9$ times faster than that with {\sc DenseAccum} over both LJ and FRIEND datasets.
	This is because the input graph data is sparse, i.e., most vertices have few edges. In each iteration, every slave node only updates the PageRank values for a small number of destination vertices.
	{\sc SparseAccum} reduces the network cost by transferring a small number of non-zero PageRank updates and hence achieves better performance than {\sc DenseAccum}. 
}

\subsection{Fine-grained DSM vs Coarse-grained DSM}
\begin{figure}[t]
	\centering	
	\subfigure[\small NETFLIX]{
		\label{fig:nmf_fine:netflix}
		\hspace{-2mm}\epsfig{height=2.8cm,width=0.5\linewidth,file=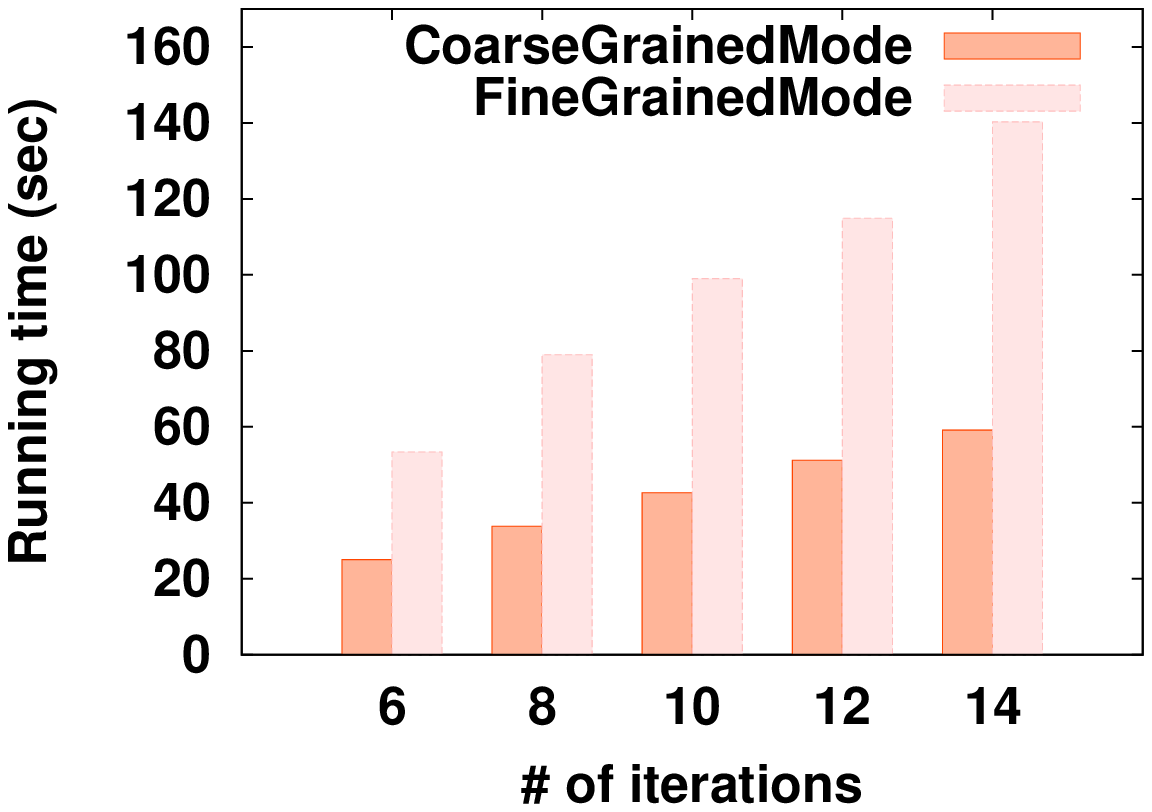}}
	\hspace{-1mm}\subfigure[\small NMFS]{
		\label{fig:nmf_fine:syn}
		\hspace{-2mm}\epsfig{height=2.8cm,width=0.5\linewidth,file=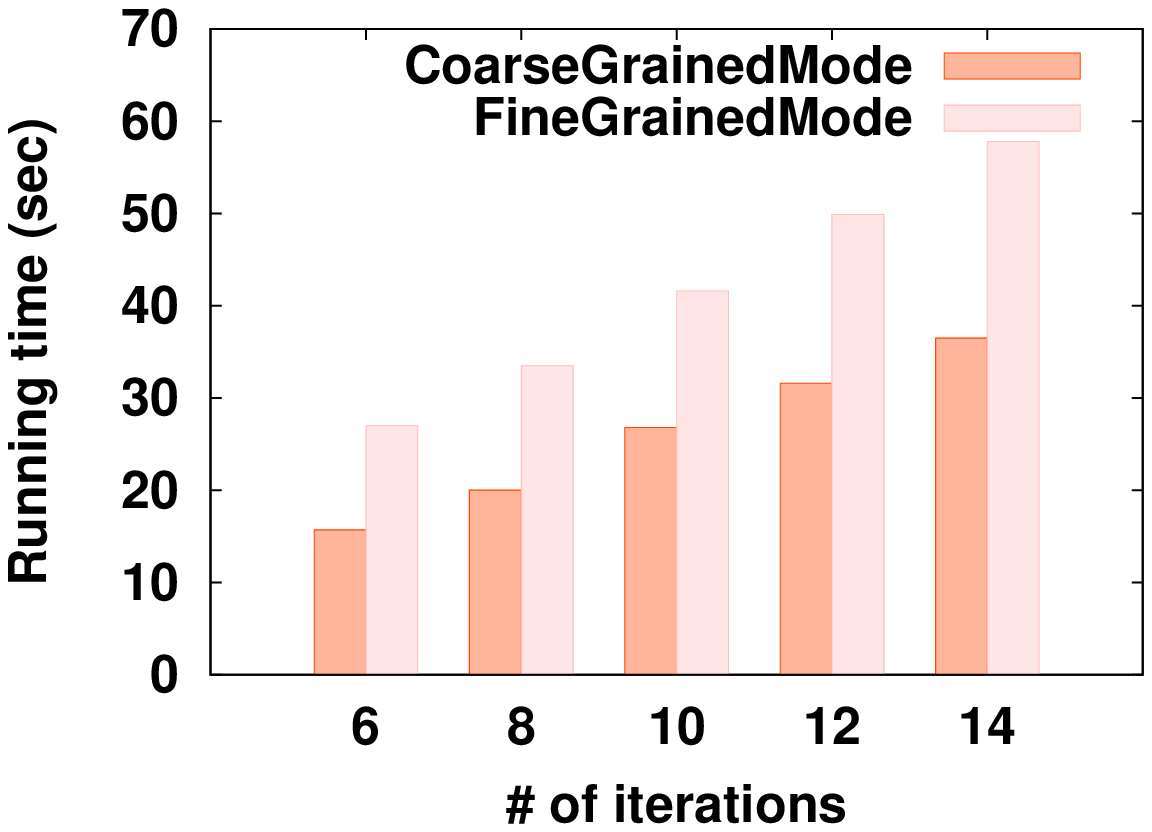}}
	\vspace{-.2in}
	\caption{\small Results on different DSM modes (NMF)}
	\label{fig:nmf_fine}
\end{figure}

We first evaluate the performance of \bird using two different modes of DSM: fine-grained mode and coarse-grained mode (see details in Section~\ref{sec:kv}).
Figure~\ref{fig:nmf_fine} shows the results of two modes for NMF.
Coarse-grained mode outperforms fine-grained mode over two datasets.
On average, \bird with coarse-grained DSM runs $2.7$ and $1.6$ times faster than that with fine-grained DSM on NETFLIX and NMFS datasets, respectively.
In NMF, we need to retrieve the factorized matrices from DSM in each iteration. Bulk loading of shared data reduces the number of data access requests and makes better use of the bandwidth of the underlying key-value store. 
Furthermore, coarse-grained DSM is more robust to the number of iterations, i.e., the increasing rate of the running time is lower than that with fine-grained DSM. 
We observe similar results on other applications and omit the figures due to redundancy.
Henceforth, all the results of \bird system are based on coarse-grained DSM. 

\eat{ 
	We now compare the performance of Birdee's coarse-grained DSM mode and fine-grained DSM mode by testing both modes in the application of NMF on both datasets. We run the same Birdee program of NMF for both DSM mode. As we can see in Figure \ref{fig:nmf_fine}, coarse-grained DSM mode has a significant advantage over fine-grained DSM mode. The application with coarse-grained DSM mode runs faster than fine-grained DSM mode application by a factor of 2.77 on synthetic data set and 3.16 on Netflix dataset. We observe that most of the use of DSM in NMF application is to operate the shared matrix Q, and coarse-grained DSM mode is more suitable for the operations on large chunks of data (like matrices and vectors), which is a common pattern in many big data processing programs, including NMF. We think the reason is that in coarse-grained DSM mode, a read or write request to the key-value store can carry multiple words, bringing a larger bandwidth of DSM.
}

\subsection{Logistic Regression}

\begin{figure}[t!]
	\vspace{-.1in}
	\centering
	\subfigure[\small \# of iterations]{
		\label{fig:lr_gene:itr}
		\hspace{-2mm}\epsfig{height=2.8cm,width=0.5\linewidth,file=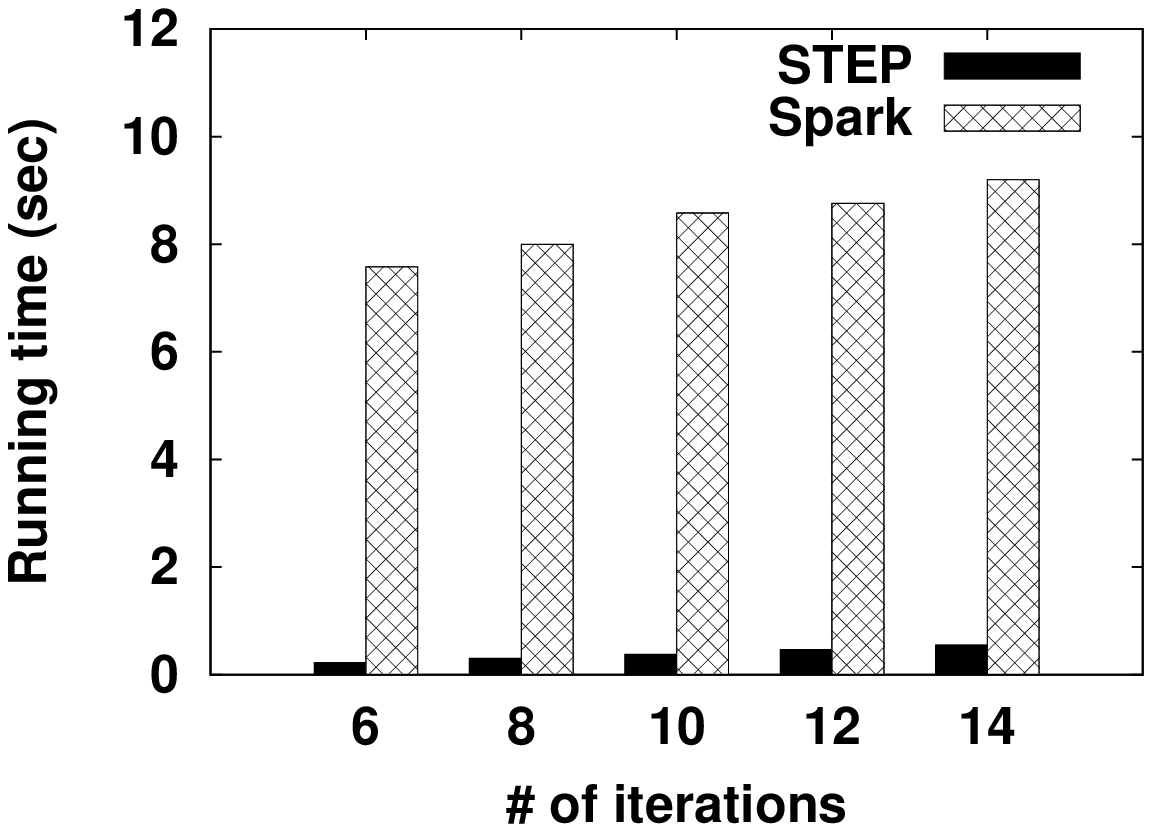}}
	\hspace{-2mm}\subfigure[\small \# of nodes]{
		\label{fig:lr_gene:vm}
		\epsfig{height=2.8cm,width=0.5\linewidth,file=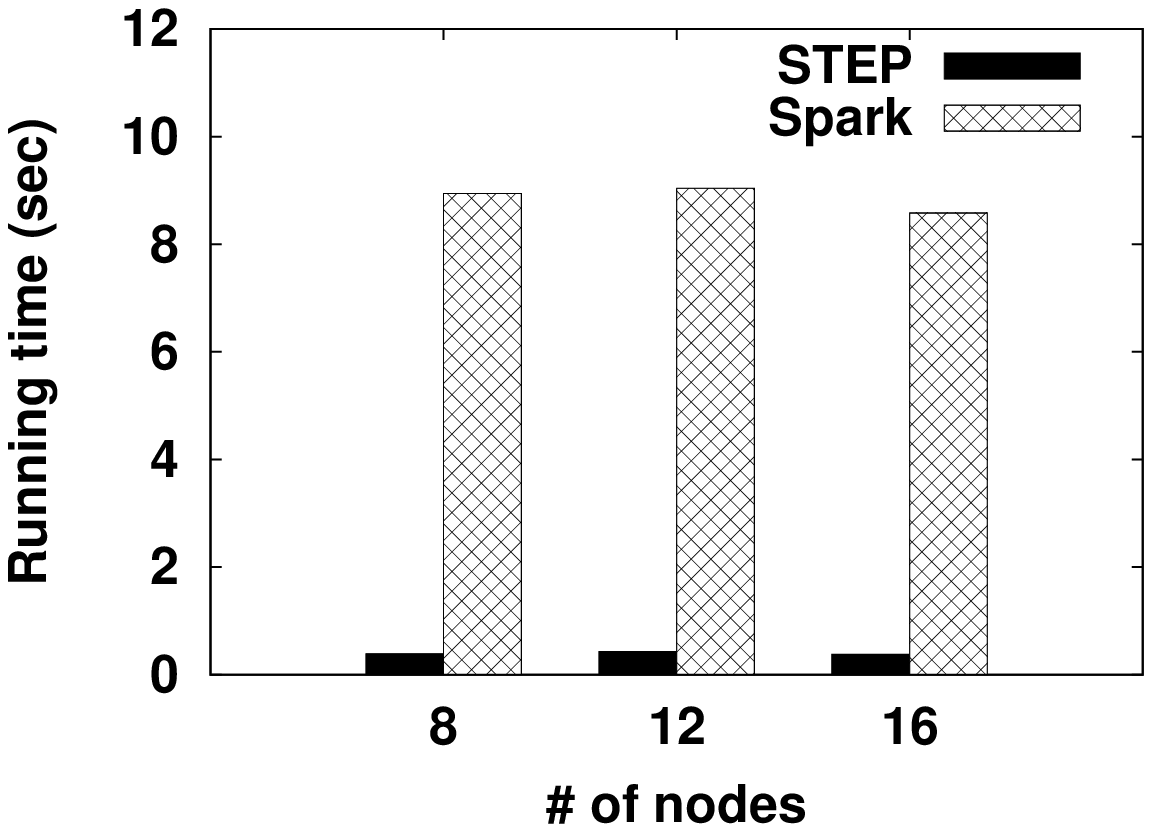}}
	\vspace{-.2in}
	\caption{\small Logistic regression results on 
		GENE dataset}
	\label{fig:lr_gene}
%
	\centering
	\subfigure[\small \# of Iterations]{
		\label{fig:lr_syn:itr}
		\hspace{-2mm}\epsfig{height=2.8cm,width=0.5\linewidth,file=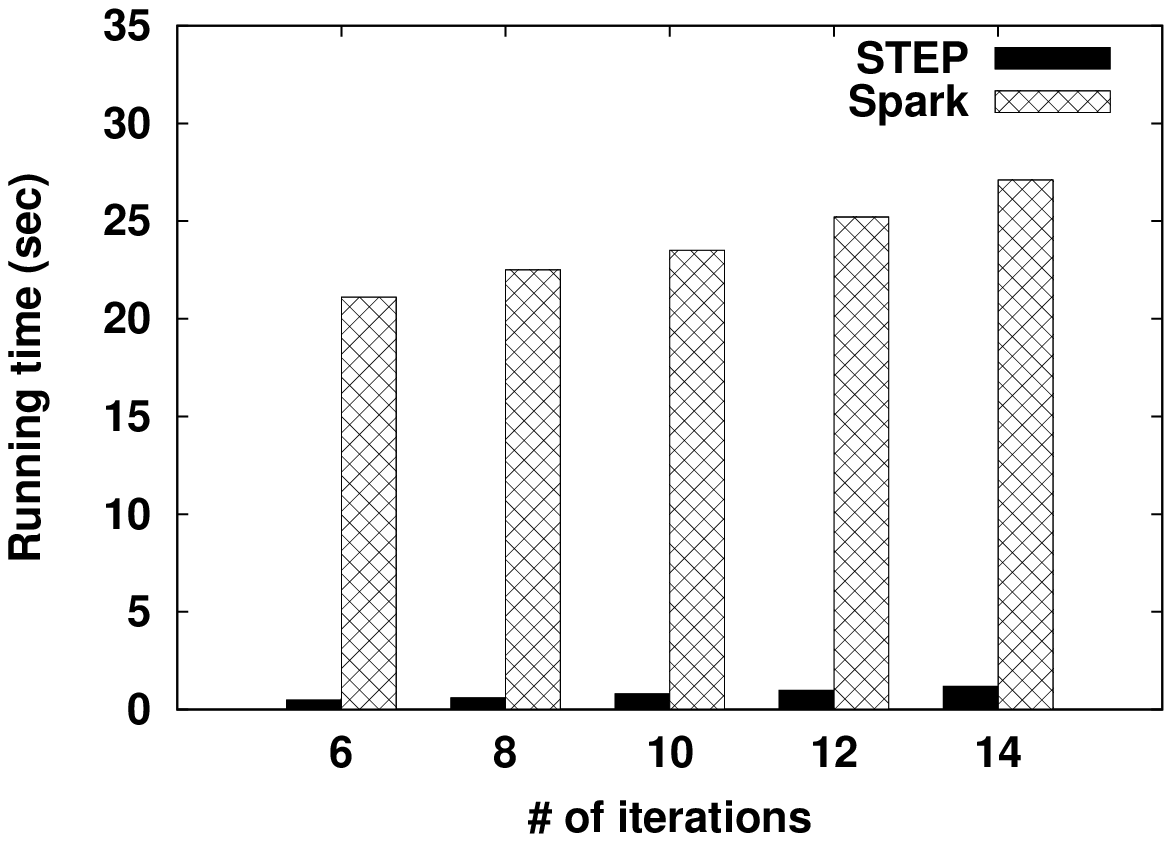}}
	\hspace{-2mm}\subfigure[\small \# of nodes]{
		\label{fig:lr_syn:vm}
		\epsfig{height=2.8cm,width=0.5\linewidth,file=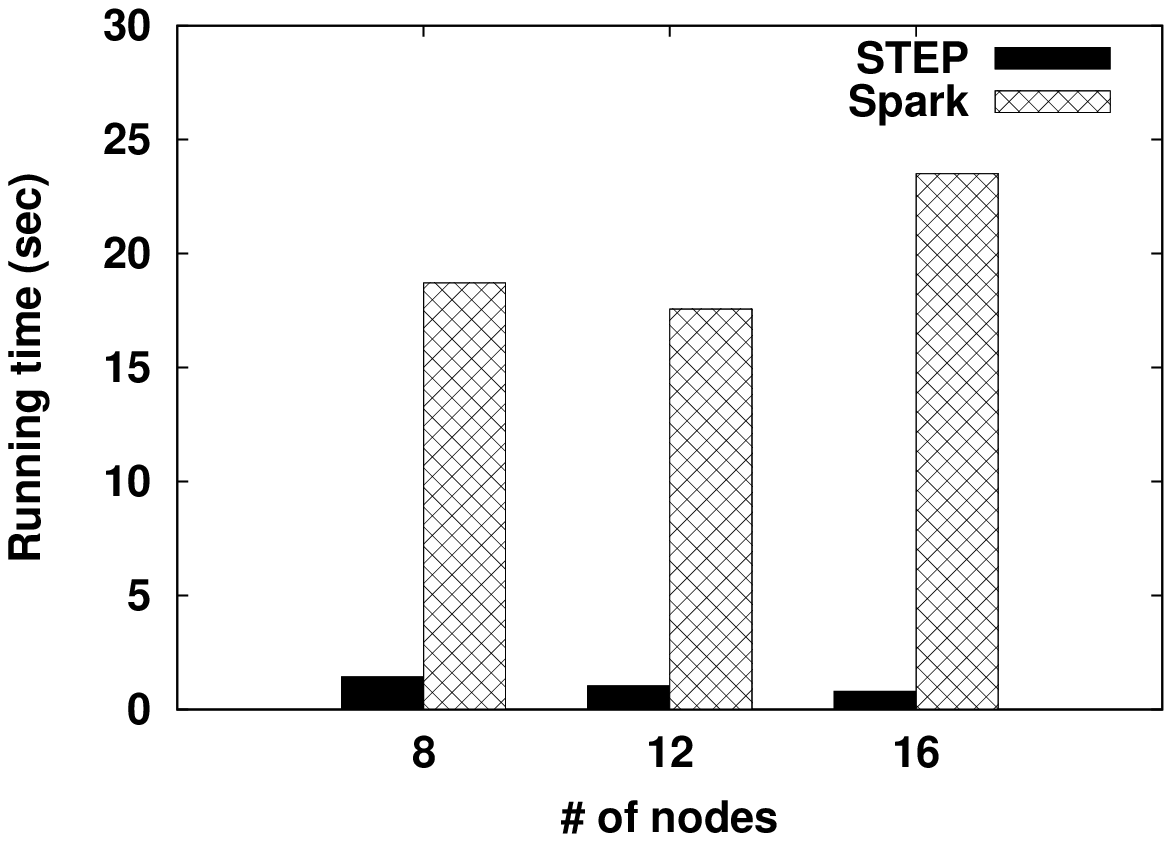}}
	\vspace{-.2in}
	\caption{\small Logistic regression results on LRS dataset}
	\label{fig:lr_syn}
	\vspace{-.1in}
\end{figure}

\begin{figure*}[th!]
	\centering
	\subfigure[\small \# of clusters (K)]{
		\label{fig:km_syn:k}
		\epsfig{height=3.2cm,width=0.28\linewidth,file=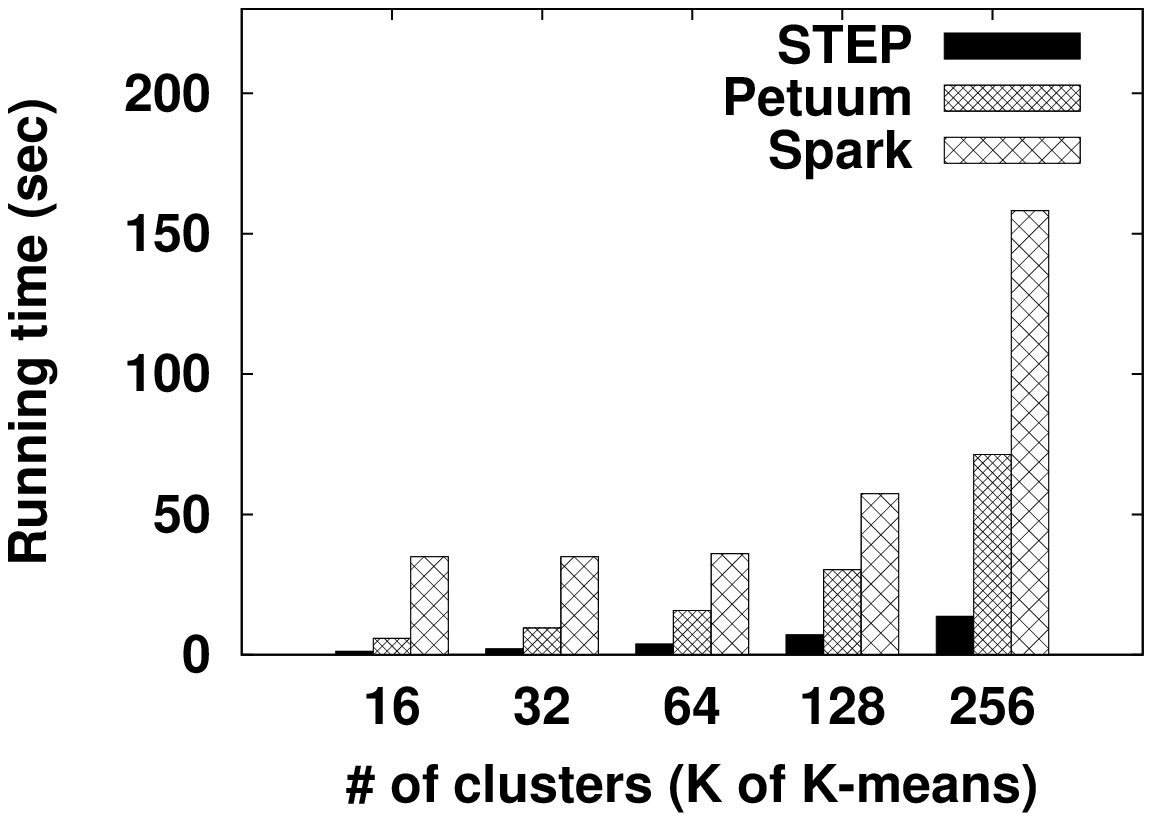}}
	\quad
	\subfigure[\small \# of iterations]{
		\label{fig:km_syn:itr}
		\epsfig{height=3.2cm,width=0.28\linewidth,file=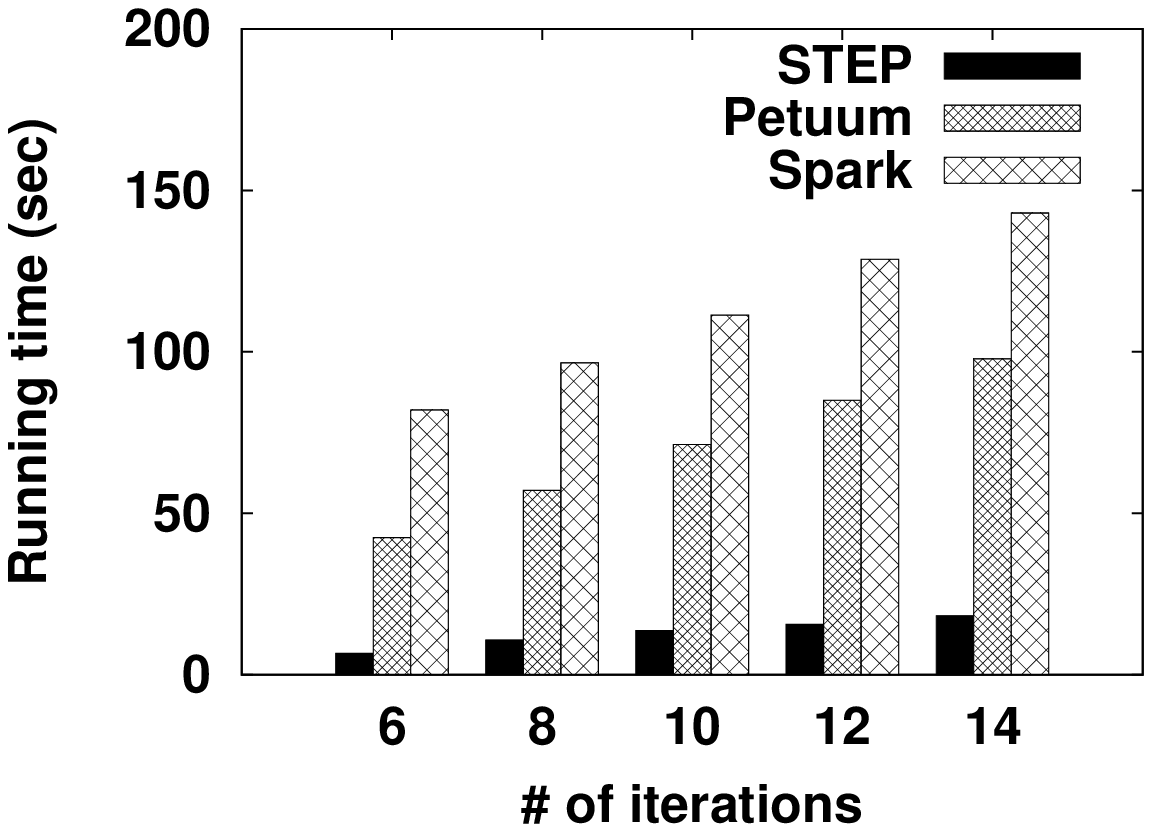}}
	\quad
	\subfigure[\small \# of nodes]{
		\label{fig:km_syn:vm}
		\epsfig{height=3.2cm,width=0.28\linewidth,file=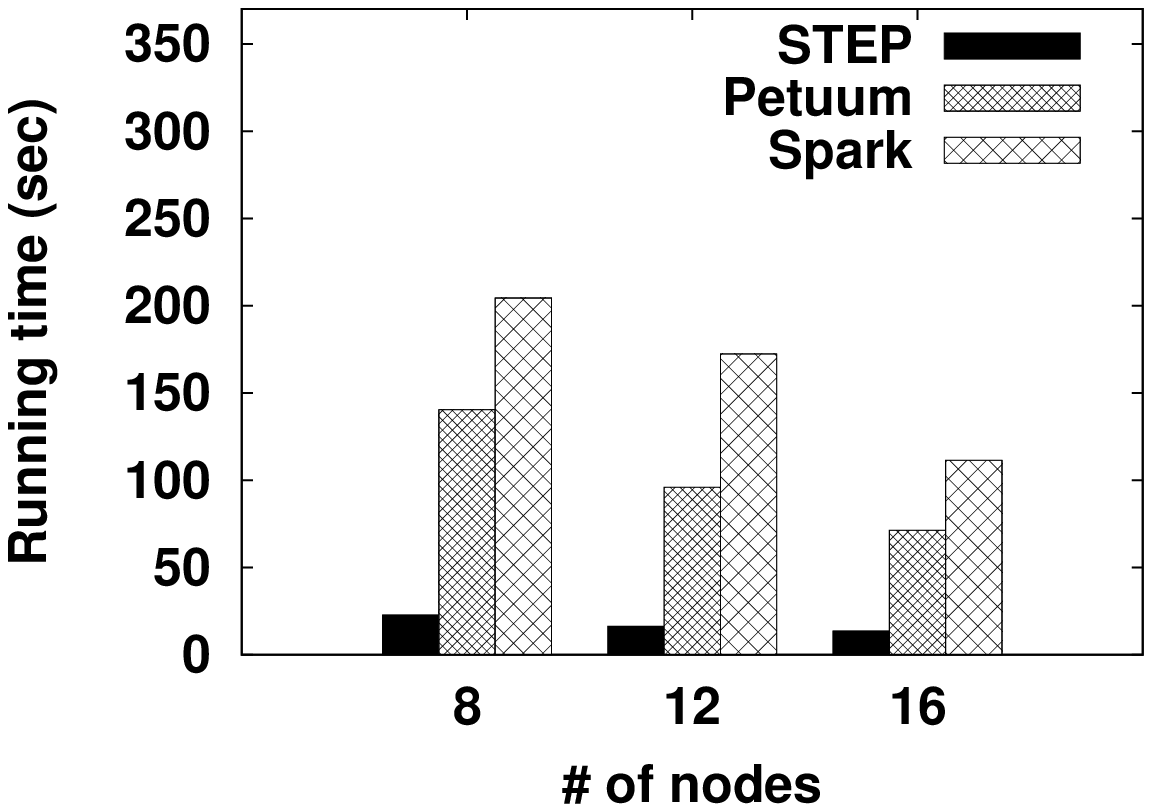}}
	\vspace{-.2in}
	\caption{\small K-means results on KMS dataset}
	\label{fig:km_syn}
	
	\centering
	\vspace{-.1in}
	\subfigure[\small K (\# of clusters)]{
		\label{fig:km_forest:k}
		\epsfig{height=3.3cm,width=0.28\linewidth,file=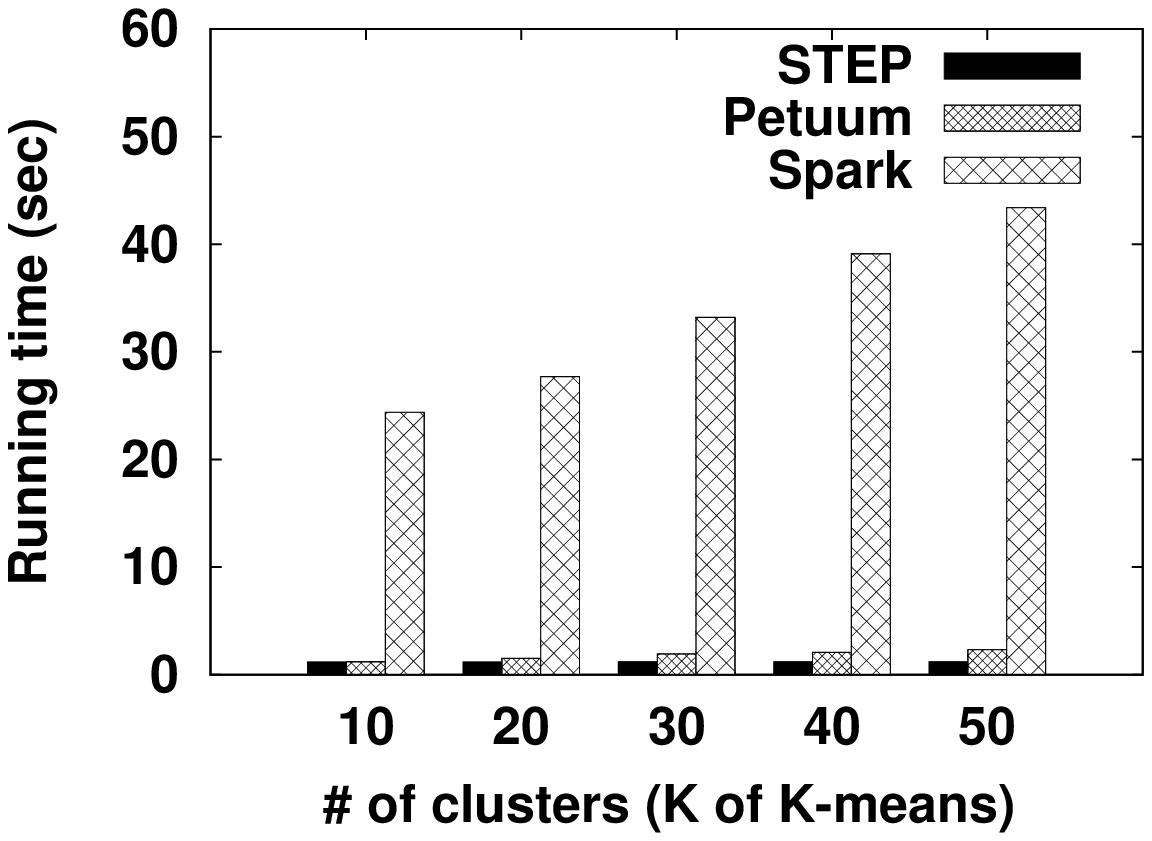}}
	\quad
	\subfigure[\small \# of iterations]{
		\label{fig:km_forest:itr}
		\epsfig{height=3.3cm,width=0.28\linewidth,file=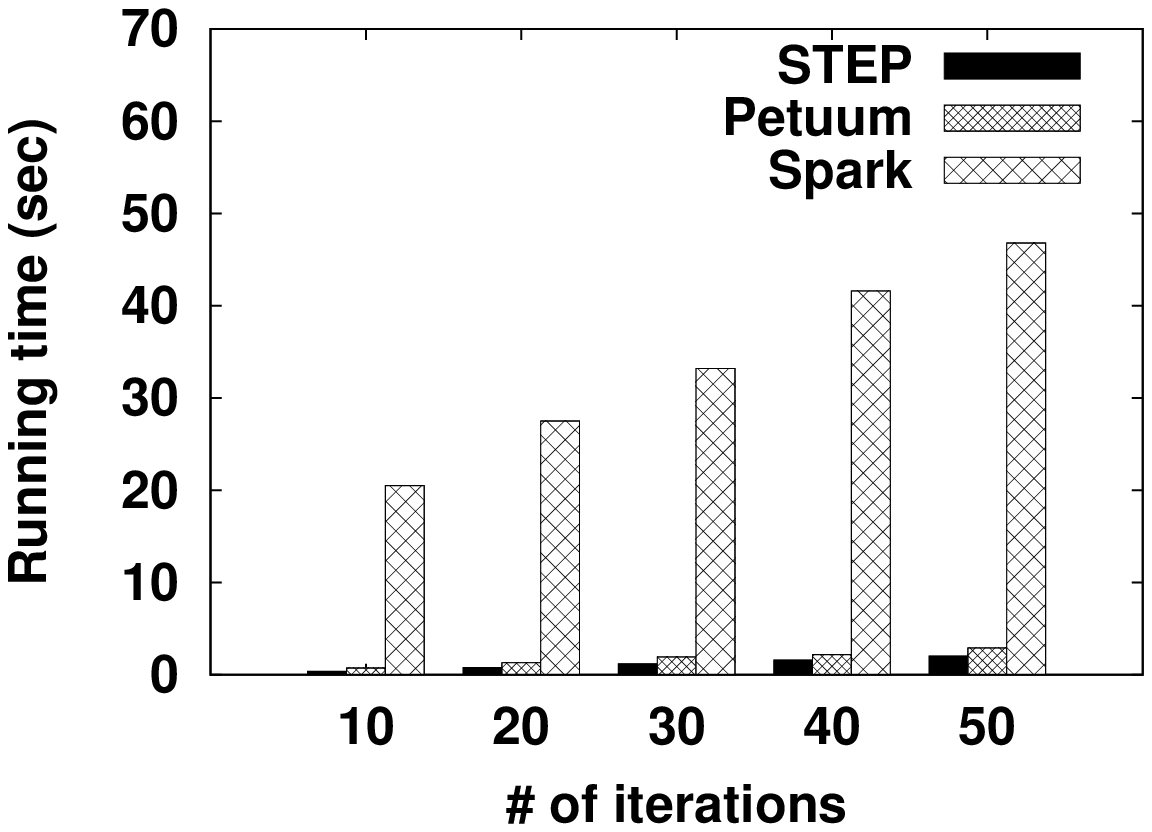}}
	\quad
	\subfigure[\small \# of nodes]{
		\label{fig:km_forest:vm}
		\epsfig{height=3.3cm,width=0.28\linewidth,file=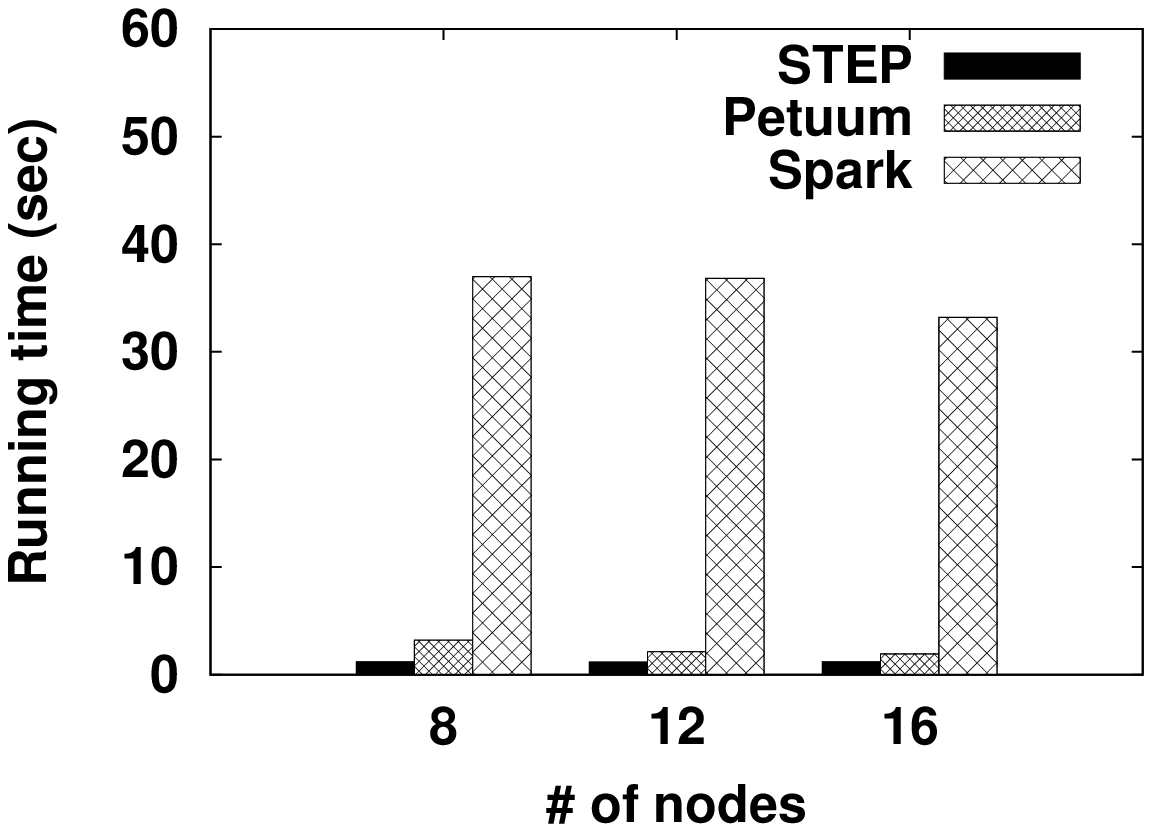}}
	\vspace{-.2in}
	\caption{\small K-means results on FOREST dataset}
	\label{fig:km_forest}
	\vspace{-.1in}
\end{figure*}

The algorithm for logistic regression has been discussed in Section~\ref{subsec:prog:lr}. 
Figure~\ref{fig:lr_gene:itr} shows the running time of different systems with various number of iterations on GENE dataset. 
\bird outperforms Spark over all the iteration numbers. 
On average, it runs $21.6$ times faster than Spark.
Both \bird and Spark requires longer running time with the increase in the number of iterations. 
In Spark, partially updated regression parameters are first forwarded to the master who will aggregate all the updates and broadcast the result to all the workers. In contrast, \bird uses accumulator to perform aggregation in parallel and leverages DSM to keep the shared parameters, both of which reduce the computation and communication burden on the master.
%
Figure~\ref{fig:lr_gene:vm} provides the results of varying the number of compute nodes over GENE dataset.
\bird runs much faster than Spark over all the node numbers.
We observe both systems do not benefit significantly from using more nodes. The reason may be that the performance gain of adding more nodes is suppressed by the increase of communication cost.

Figure~\ref{fig:lr_syn:itr} shows the logistic regression results on LRS dataset.
Both \bird and Spark require longer running time as the number of iterations increases. 
Similar to GENE dataset, \bird runs much faster than Spark over all the iterations, i.e., $29$ times faster than Spark on average. 
%
Figure~\ref{fig:lr_syn:vm} provides the running time on LRS dataset by varying the number of compute nodes.
When the number of nodes increases from 8 to 16, the running time of \bird is decreased linearly. Spark achieves the best performance with 12 nodes and the performance is decreased by using 16 nodes. This is because LRS is a large high-dimensional dataset. When using more nodes, Spark master has to aggregate more updated parameters and broadcast the results to more workers, which incurs very high communication cost and compromises the gain in computation time.


\eat{
	that Birdee is on average 5.9 times faster than Spark in the task of logistic regression on E-TABM-185 dataset. In addition, Birdee's performance gets a 17\% gain from doubling the number of machines from 8 to 16. In contrast, Spark gets no almost benefits from adding more machines. The reason for the result above may attribute to Birdee's more flexible memory abstraction. Birdee programs can re-use the memory on both shared and local memory space in each iteration, making them more suitable for iterative tasks, which leads to a performance advantage over Spark, while Spark's memory abstraction RDD is less flexible. RDDs created in the current iteration have to be disposed in the next iteration for it is immutable. This brings more pressure on Spark's garbage collector, leaving an affect on its performance. 
}

\subsection{K-means}
\label{k-means}


We first study the effects of varying the number of clusters on K-means clustering using KMS dataset. 
Figure~\ref{fig:km_syn:k} shows the running time for various K. 
Spark requires the longest running time over all the values of K, and the disadvantage becomes more significant when K increases. For K=256, Spark is over $10$ times slower than \bird. 
We attribute the poor performance to the deficiency of the programming language of Spark and 
the master node who is responsible for computing and broadcasting K centers in each iteration.
%
\bird performs better than Petuum for all values of K. For K=256, \bird runs $5.4$ times faster than Petuum. 
The reason is that Petuum's K-means application transfers K center vectors by sparse vectors, which may result in more communications than using dense vectors in some datasets. While the accumulator in \bird can automatically detect the sparsity of the vector, choosing more efficient  way to transfer the shared data.
The running time of all three systems increases when K becomes larger, due to the higher computation and communication cost.

Figure~\ref{fig:km_syn:itr} provides the running time on KMS dataset when the number of iterations varies from $6$ to $14$. 
All the systems require longer running time as the number of iterations increases. 
On average, \bird runs $5.4$ and $8.6$ times faster than Petuum and Spark, respectively. The advantage of \bird is consistent over all the iteration numbers.

Figure~\ref{fig:km_syn:vm} reports the results with various node numbers.
\bird is more efficient than Petuum and Spark when the number of nodes increases from 8 to 16. 
With $8$ compute nodes, \bird is $2.7$ and $12$ times faster than Petuum and Spark, respectively. 
When we double the number of nodes from $8$ to $16$, Petuum achieves the highest speedup of $1.96$, followed by a speedup of $1.83$ by Spark. \bird achieves $1.67$ speedup, which is lower than Petuum and Spark. 
This is because the computation in \bird is already efficient with a small number of compute nodes, and the performance does not benefit too much from using more nodes.


Figure~\ref{fig:km_forest} provides K-means results on FOREST dataset.
\bird outperforms Spark and Petuum over all the values of K.
Both \bird and Petuum takes longer running time when K becomes larger.
This is because larger values of K require more distance computations and comparisons.
We observe similar results {with the increase of iteration numbers} on FOREST dataset, where \bird runs $1.5$ times faster than Petuum and $28$ times faster than Spark on average.

\subsection{NMF}
\label{NMF}

\begin{figure*}[t!]
	%
	\centering
	\subfigure[\small \# data points]{
		\label{fig:nmf_syn:rank}
		\epsfig{height=3.2cm,width=0.28\linewidth,file=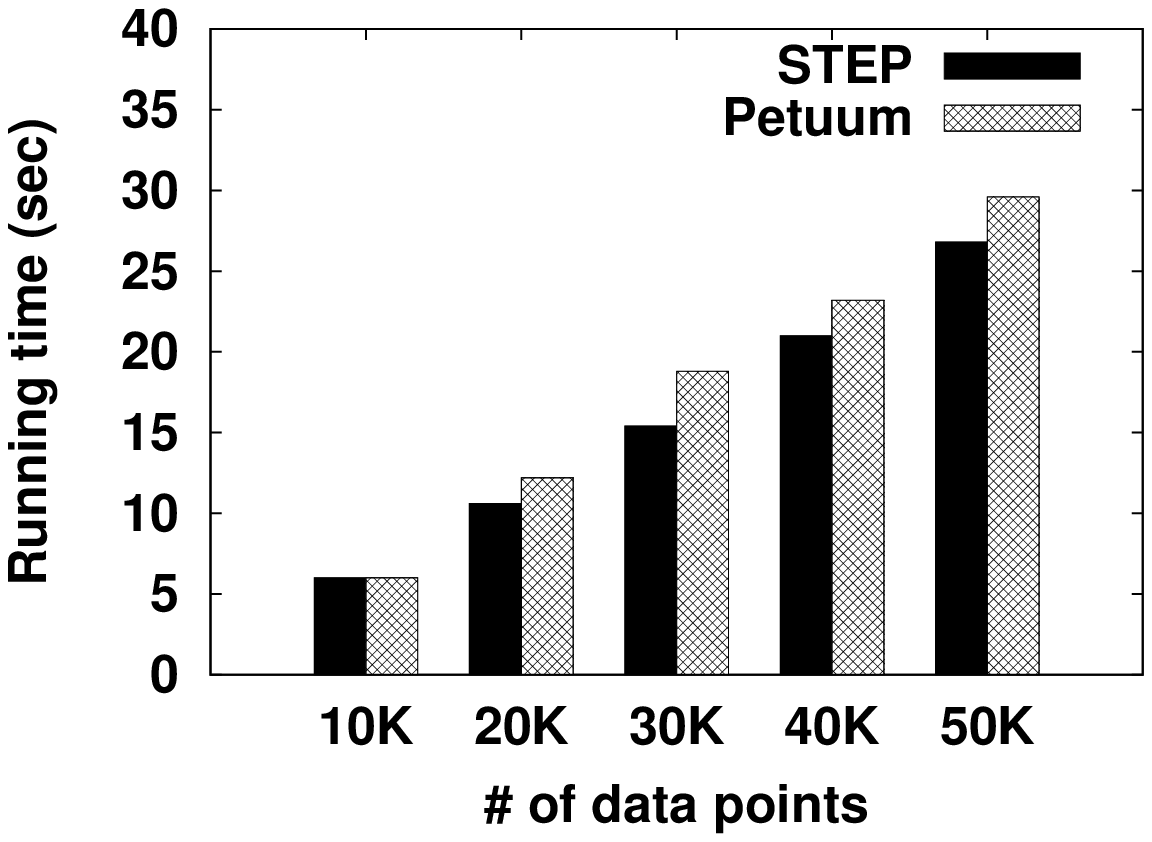}}
	\quad
	\subfigure[\small \# of iterations]{
		\label{fig:nmf_syn:itr}
		\epsfig{height=3.2cm,width=0.28\linewidth,file=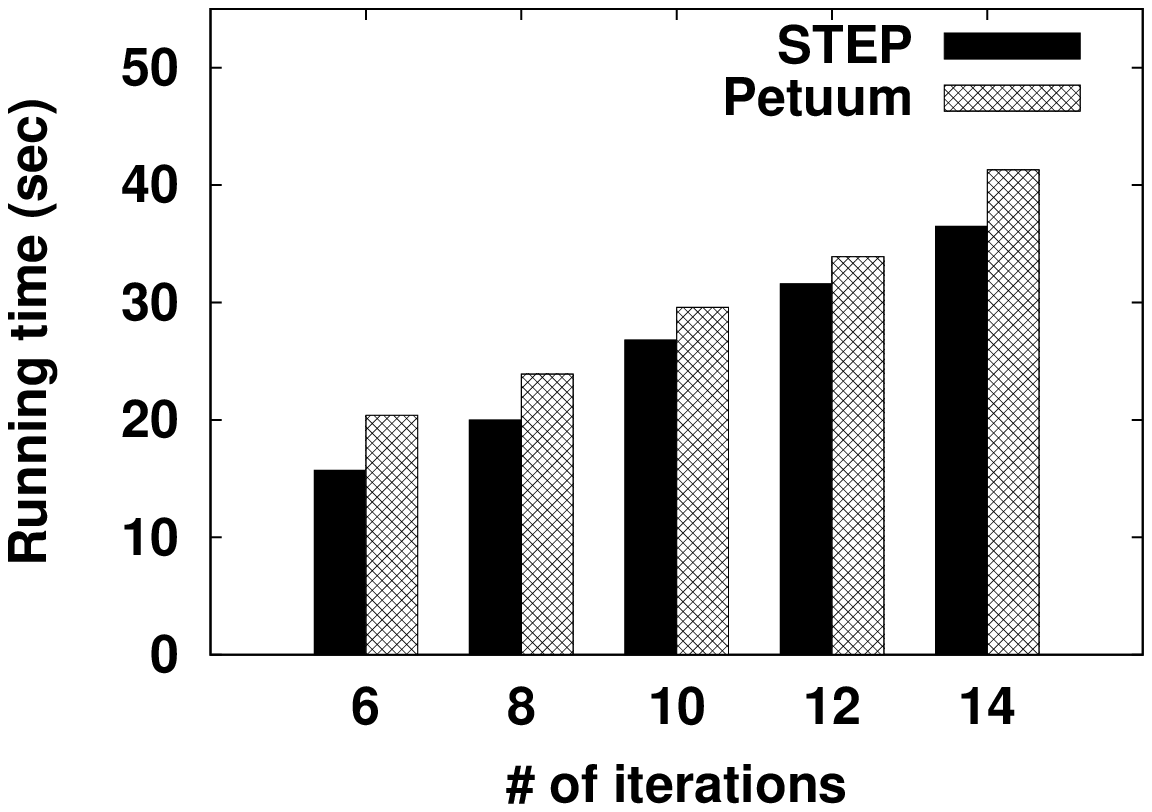}}
	\quad
	\subfigure[\small \# of nodes]{
		\label{fig:nmf_syn:vm}
		\epsfig{height=3.2cm,width=0.28\linewidth,file=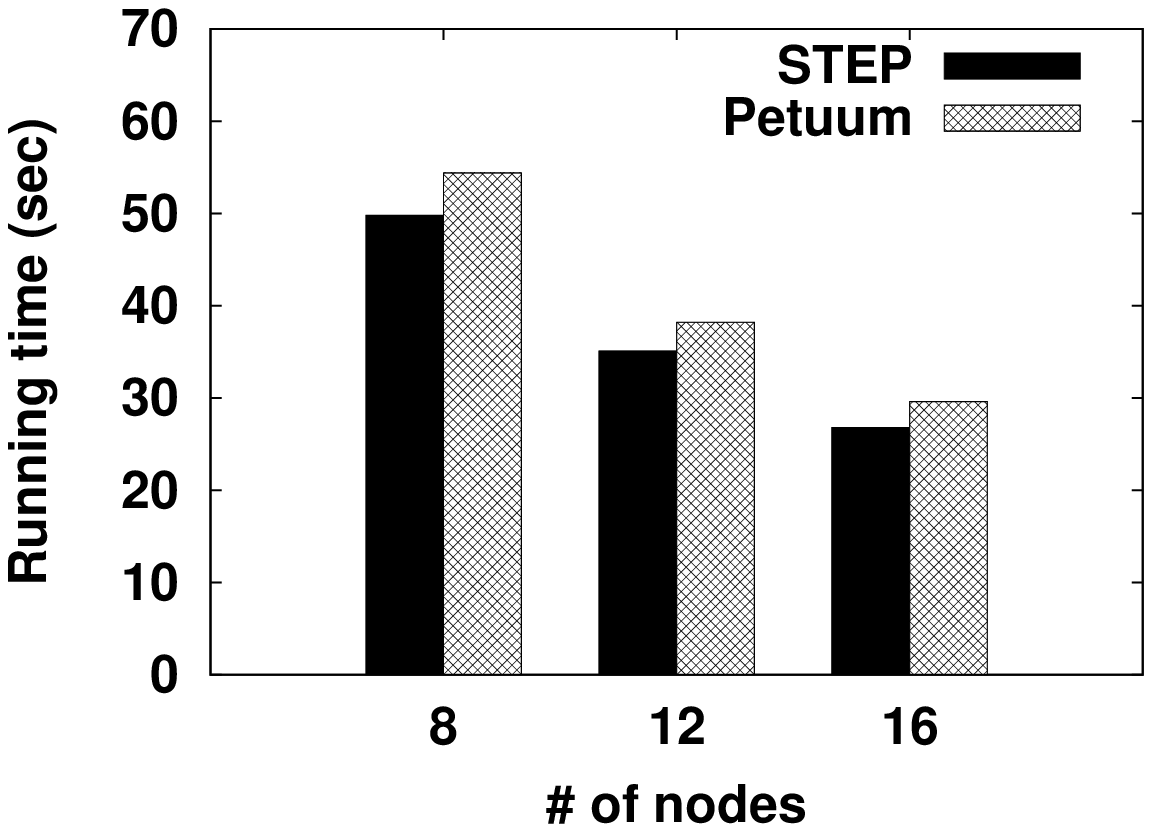}}
	\vspace{-.2in}
	\caption{\small NMF results on NMFS dataset}
	\label{fig:nmf_syn}
	
	\centering
	\subfigure[\small Factorization rank]{
		\label{fig:nmf_netflix:rank}
		\epsfig{height=3.3cm,width=0.28\linewidth,file=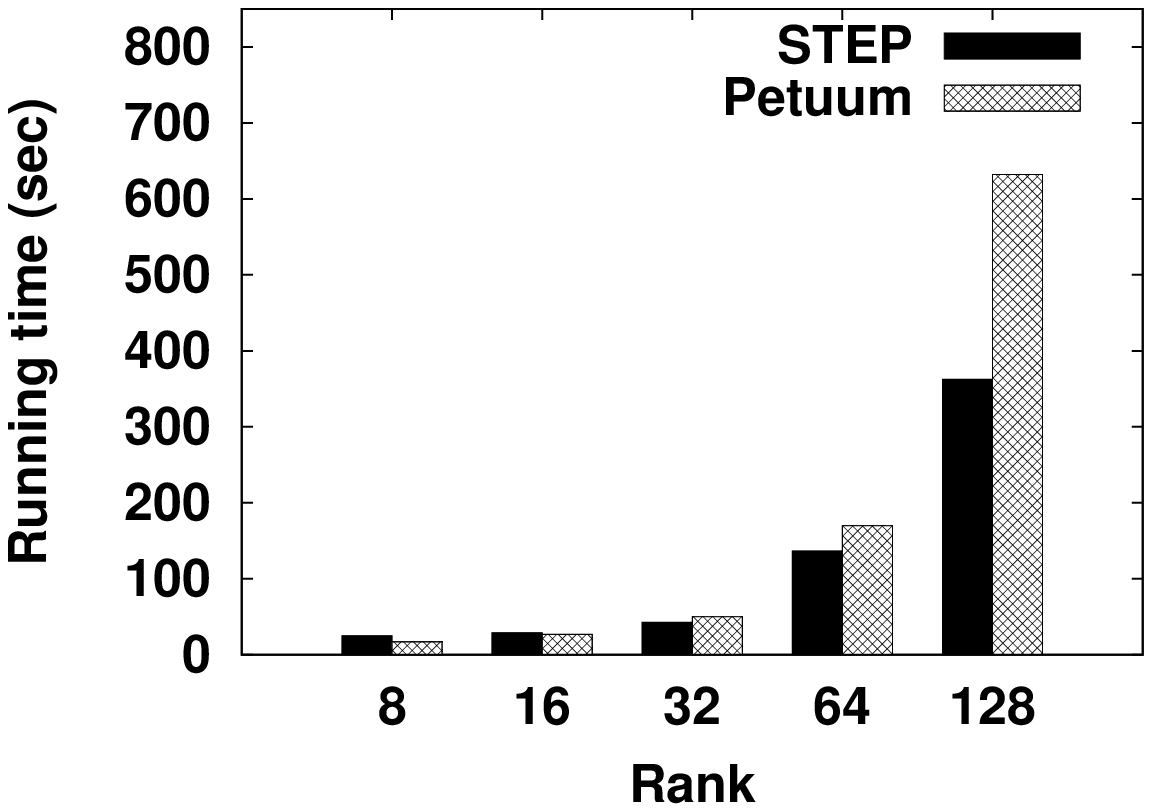}}
	\quad
	\subfigure[\small \# of iterations]{
		\label{fig:nmf_netflix:itr}
		\epsfig{height=3.3cm,width=0.28\linewidth,file=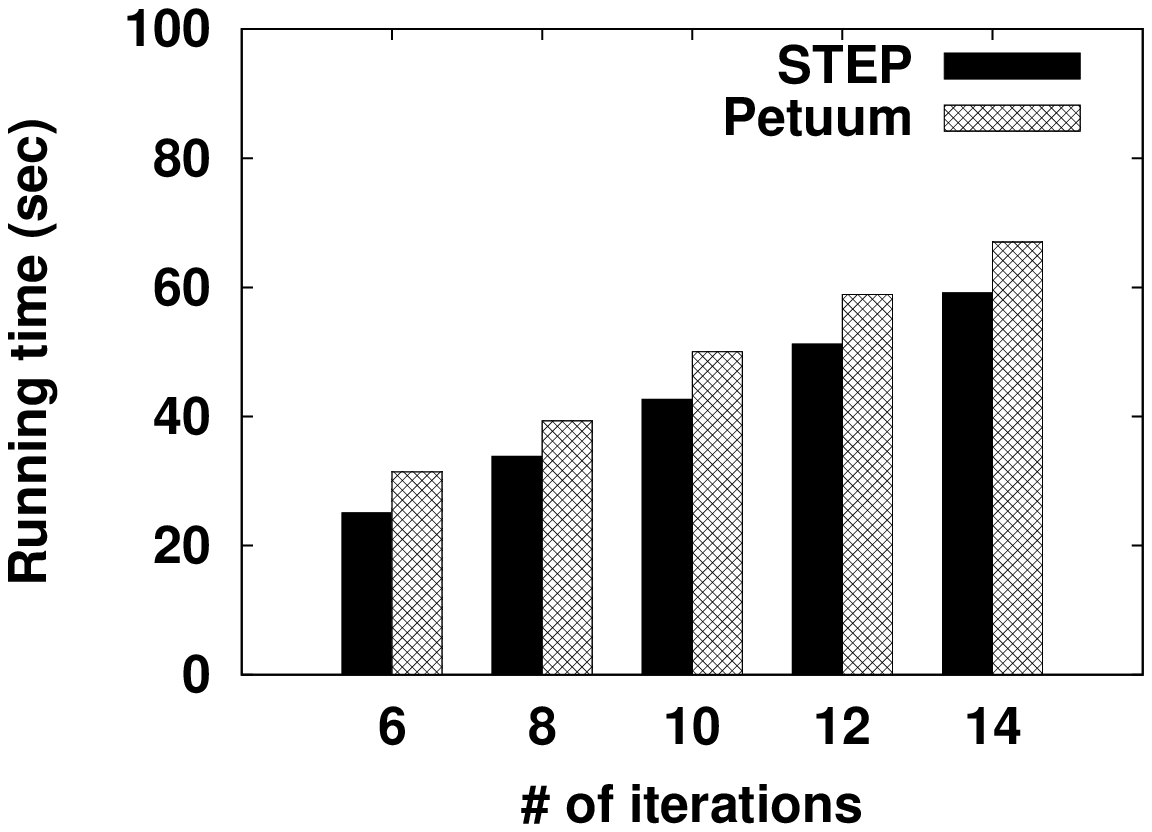}}
	\quad
	\subfigure[\small \# of nodes]{
		\label{fig:nmf_netflix:vm}
		\epsfig{height=3.3cm,width=0.28\linewidth,file=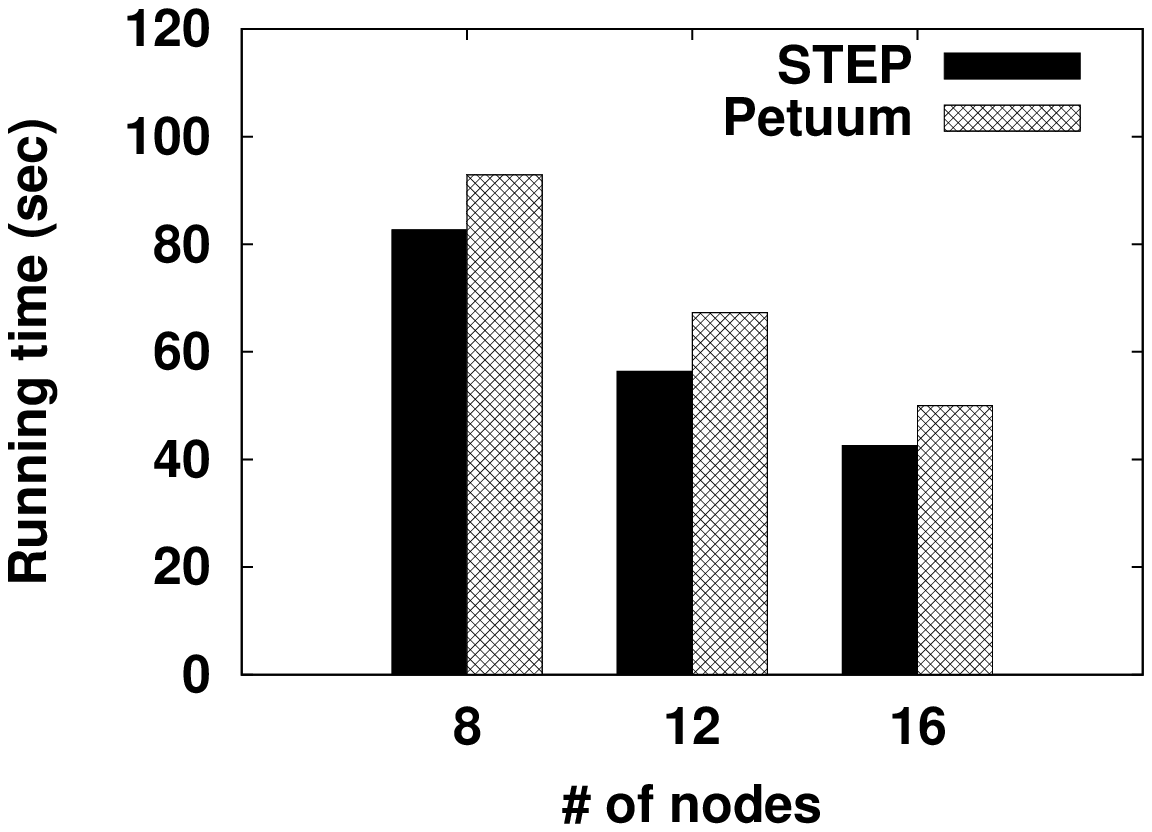}}
	\vspace{-.2in}
	\caption{\small NMF results on NETFLIX dataset}
	\label{fig:nmf_netflix}
	
	\centering
	\subfigure[\small \# of iterations (LJ)]{
		\label{fig:pr_lj:itr}
		\hspace{-1mm}\epsfig{height=2.8cm,width=0.24\linewidth,file=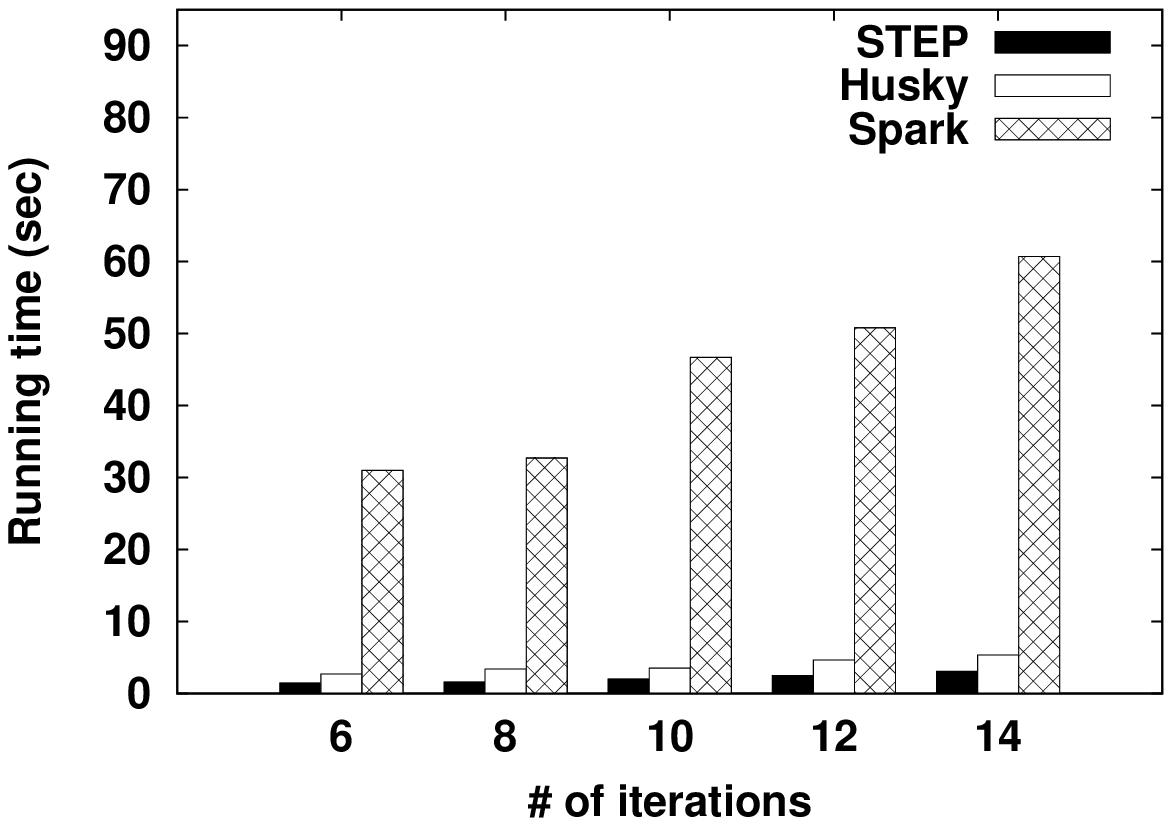}}
	\subfigure[\small \# of nodes (LJ)]{
		\label{fig:pr_lj:vm}
		\epsfig{height=2.8cm,width=0.24\linewidth,file=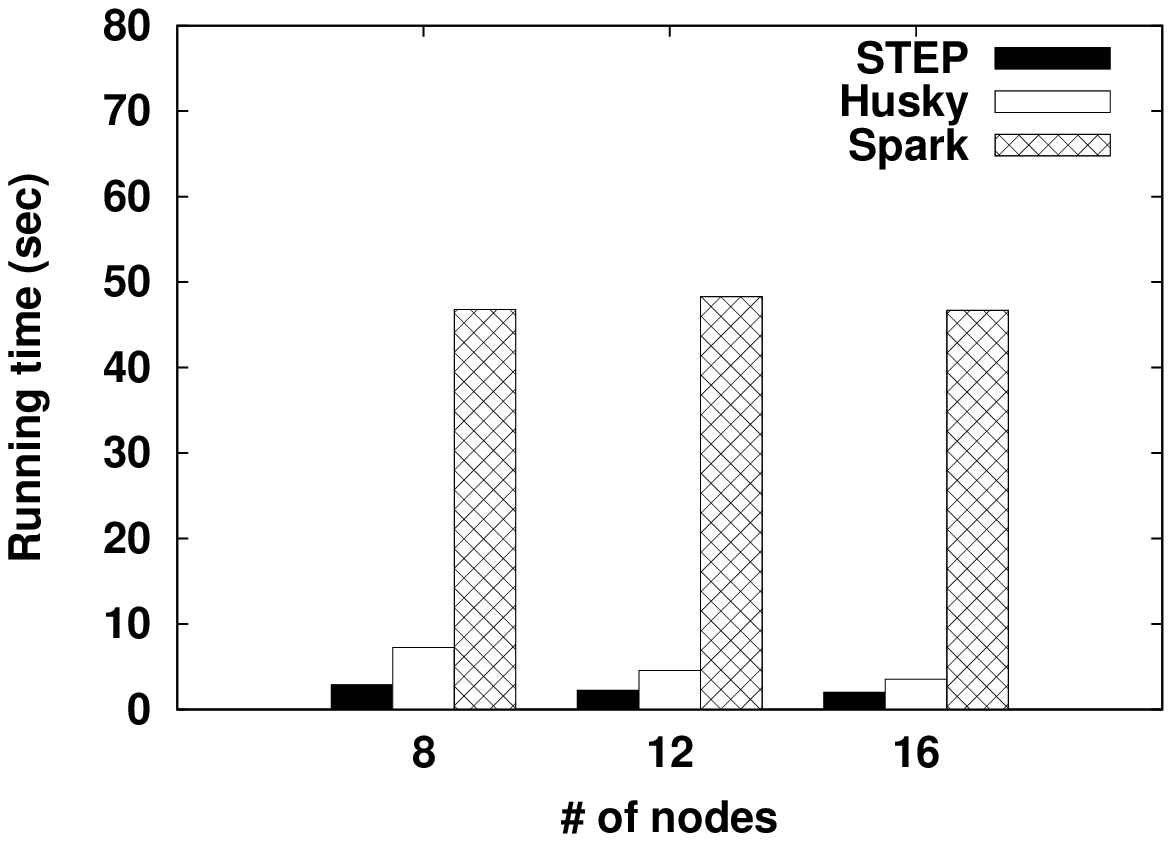}}
	\subfigure[\small \# of iterations (FRIEND)]{
		\label{fig:pr_friend_itr}
		\epsfig{height=2.8cm,width=0.24\linewidth,file=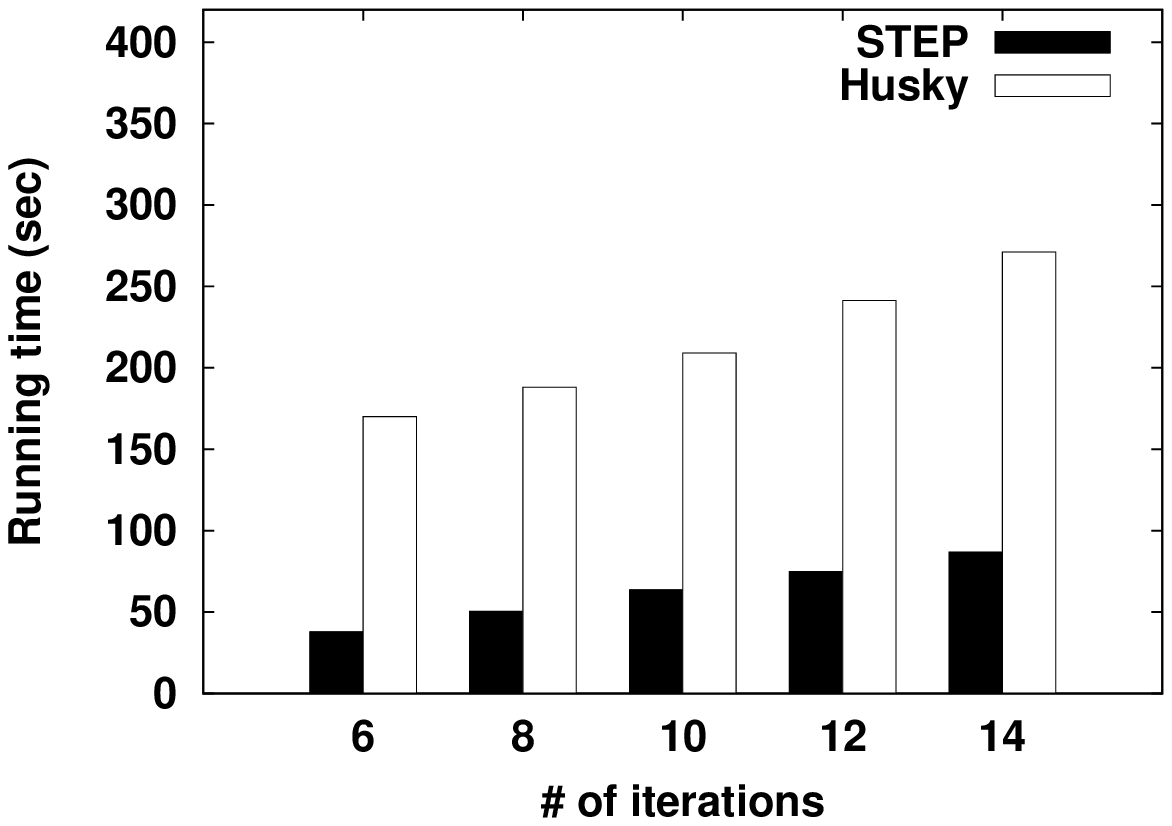}}
	\subfigure[\small \# of nodes (FRIEND)]{
		\label{fig:pr_friend_vm}
		\epsfig{height=2.8cm,width=0.24\linewidth,file=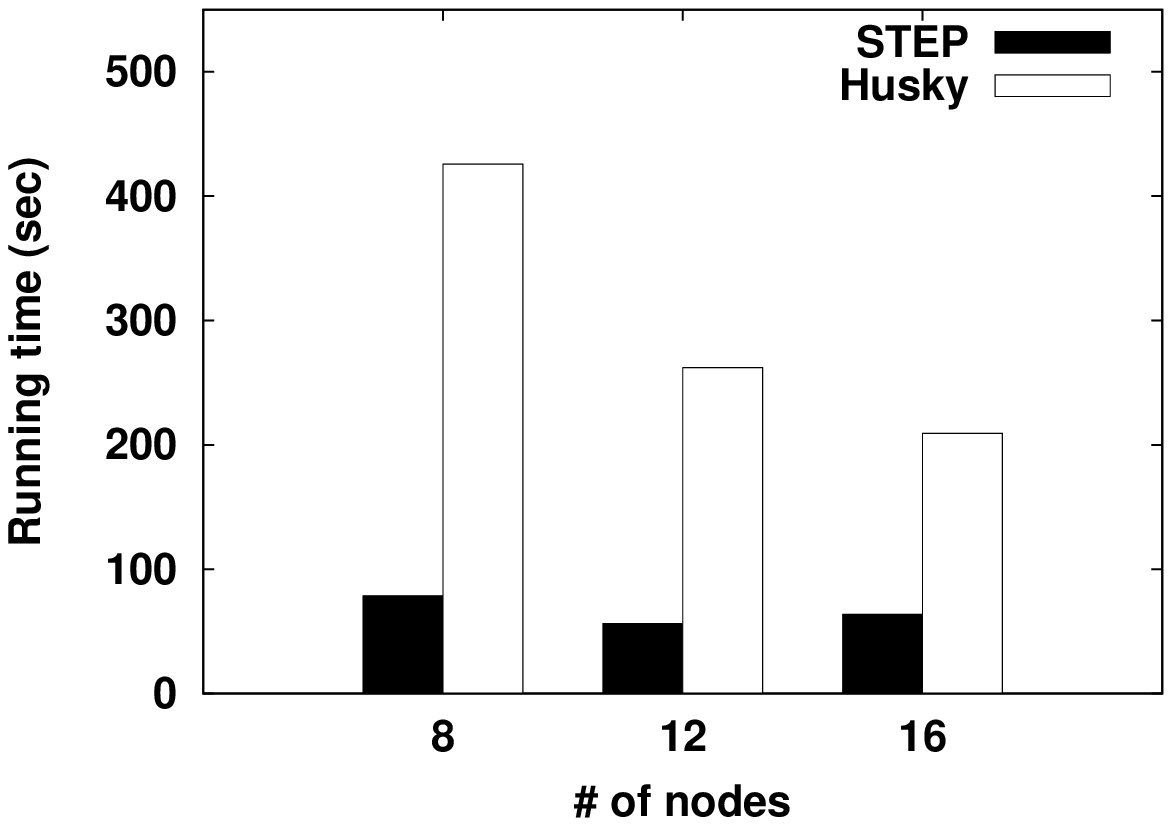}}	
	\vspace{-.2in}
	\caption{\small PageRank results}
	\label{fig:pr}
	\vspace{-.1in}
\end{figure*}

Given a factorization rank $k$, NMF tries to factorize an $n\times m$ input matrix R into two matrices P ($n\times k$) and Q ($k\times m$) so that the loss function $L=\frac{\Vert{R-PQ}\Vert^2}{n}$ is minimized.
Both \bird and Petuum adopt the SGD algorithm that learns two matrices P and Q iteratively.

Figure~\ref{fig:nmf_syn} shows the running time of \bird and Petuum on NMFS dataset.
We first evaluate the effects of varying the number of data points, i.e., data size.
We generated 500K data points for NMFS and allow the system to load a subset of data to get a smaller data size. 
Figure~\ref{fig:nmf_syn:rank} provides the running time. 
When the number of data points increases, both \bird and Petuum require longer running time to learn P and Q.
This is because of the longer data loading time and more complex training process. 
\bird outperforms Petuum over all the data point numbers, i.e., $1.1$ times faster than Petuum on average. 
We observe the computation time in \bird is less than that in Petuum. This may result from the simple shared memory abstraction  and efficient shared data manipulation in \bird.
%
Figure~\ref{fig:nmf_syn:itr} shows the effects of varying the number of iterations on the performance. For both \bird and Petuum, the running time increases linearly with the number of iterations. 
On average, \bird is $1.14$ times faster than Petuum over all the iteration numbers. 
Figure~\ref{fig:nmf_syn:vm} provides the speedup by varying the number of nodes.
\bird outperforms Petuum over all the node numbers. Both systems achieve similar speedup (i.e., $1.8$) when the node number increases from 8 to 16.

Figure~\ref{fig:nmf_netflix} shows the results of NMF on NETFLIX dataset.
\bird outperforms Petuum when the factorization rank exceeds 16. 
The difference becomes larger with the increase of factorization rank.
Since NMF with a large factorization rank puts more burden on both CPU and network, this result verifies the efficiency of \bird over the workloads with high computation and communication cost.
Both \bird and Petuum require longer running time when the number of iterations increases. 
\bird runs $1.16$ times faster than Petuum over all the iteration numbers. 
We observe that Petuum requires more computation time than \bird and hence achieves better speedup by adding more nodes.

\subsection{PageRank}\label{subsec:pr}


We now present the results of PageRank computation.
Spark runs more than 2 hours over the large FRIEND dataset. We found Spark incurs huge memory consumption during iterative computation and RDDs are flushed to disk from time to time. Hence, we only report the results of \bird and Husky on FRIEND dataset.

Figure~\ref{fig:pr_lj:itr} shows the running time by varying the number of iterations on LJ dataset. On average, \bird runs $20.8$ times faster than Spark, and outperforms Husky by a factor of $1.75$. 
Husky performs vertex-to-vertex message forwarding for PageRank computation and the communication cost in each iteration is proportional to the number of edges. 
In contrast, \bird leverages accumulator to compute PageRank values in parallel where the communication cost is proportional to the number of vertices. Furthermore,
the coarse-grained DSM exhibits high performance in accessing PageRank values for all vertices.

Figure~\ref{fig:pr_lj:vm} provides the running time of different systems by varying the number of nodes on LJ dataset.
On average, \bird runs $2.1$ and $19.6$ times faster than Husky and Spark, respectively. 
\bird achieves lower speedup than Husky ($1.4$ vs $2$) when the node number increases from 8 to 16, as the computation in \bird is already efficient with 8 compute nodes. The running time of Spark does not vary too much when the number of compute nodes is doubled. While adding more nodes improves thread-level parallelism, the network communication cost is increased as a side effect.

Figure~\ref{fig:pr_friend_itr} presents the running time on FRIEND dataset.
\bird runs about $3$ times faster than Husky over all the iteration numbers. 
The running time of both systems increases linearly with the number of iterations.
Figure~\ref{fig:pr_friend_vm} shows the results of varying the number of compute nodes. The running time of \bird and Husky decreases linearly with the increase of compute nodes, which illustrates the scalability of both systems. \bird outperforms Husky over all the node numbers. We attribute the advantage of \bird to the efficient accesses and updates of PageRank values via coarse-grained DSM and accumulator. As mentioned before, the communication cost of \bird in each iteration is proportional to the number of vertices, which is much lower than the total size of messages forwarded in Husky.

\subsection{Fault Tolerance}\label{subsec:fault}

\begin{figure}[t]
	\centering	
	\subfigure[\small Running time per iteration]{
		\label{fig:fault:itr}
		\hspace{-1mm}\epsfig{height=2.6cm,width=0.48\linewidth,file=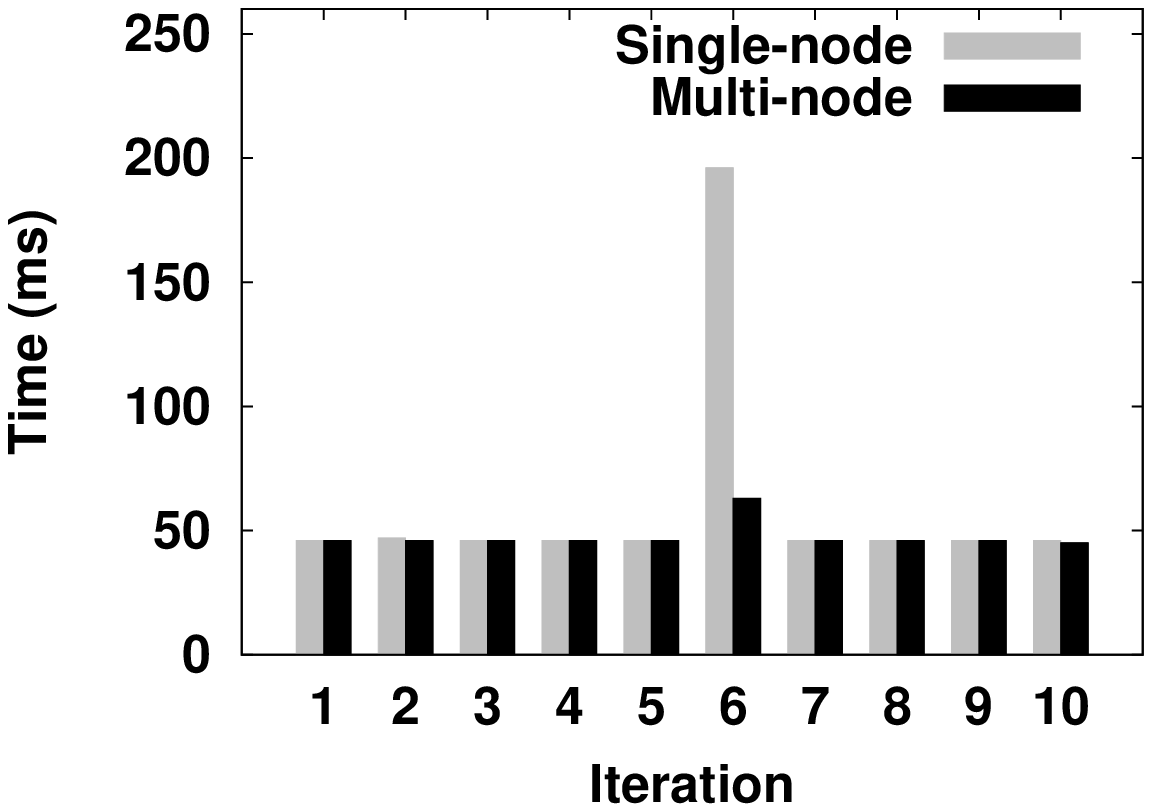}}
	\hspace{-1mm}\subfigure[\small Recovery time in iteration 6]{
		\label{fig:fault:6}
		\hspace{-1mm}\epsfig{height=2.6cm,width=0.5\linewidth,file=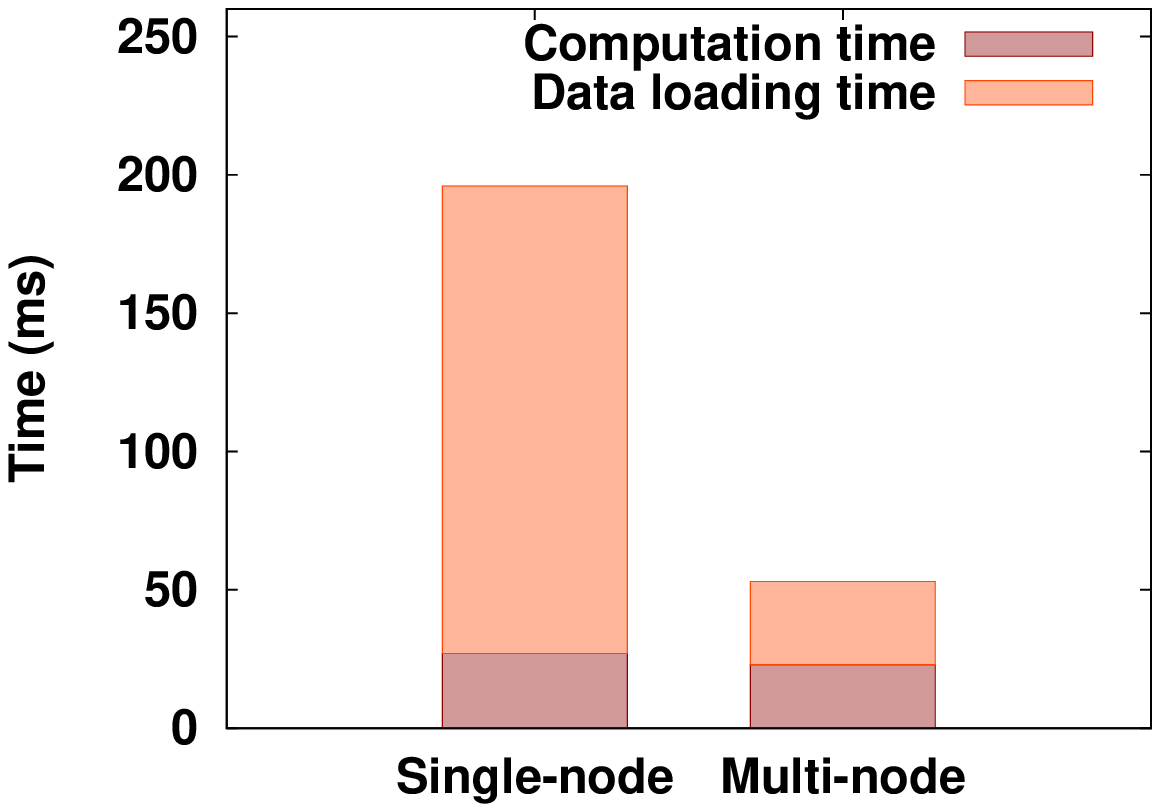}}
	\vspace{-.15in}
	\caption{\small Fault tolerance results}
	\label{fig:fault}
	\vspace{-.2in}
\end{figure}
We now evaluate the performance of fault tolerance in \bird. 
We tested two recovery methods used in \bird as follows. In \emph{single-node recovery}, when a node fails, one healthy node will take over the task for the failed node. 
Namely, all the failed working threads will be recreated in a healthy node.
In \emph{multi-node recovery}, we recreate failed threads in multiple compute nodes that perform recovery task collaboratively.
To simulate a node failure, we disconnect a slave in the 6-th iteration when running K-means over a 16-node cluster. Figure~\ref{fig:fault} shows the recovery performance of \bird. Each of the first 5 iterations takes about 46ms. In iteration 6, a node with 4 working threads fails. In \emph{single-node recovery}, \bird requires 196ms to load data using one healthy node and redo the computation for the 6-th iteration (we may require all the nodes to redo all the iterations since the latest checkpoint in other applications). Among the 196ms, we find that data loading occupies 169ms and K-means computation takes 27ms.
\emph{multi-node recovery} initializes four threads in different healthy nodes. It outperforms \emph{single-node recovery} by using only 63ms to recover from the failure, where 40ms and 23ms are used for data loading and recomputation, respectively. This is because \emph{multi-node recovery} performs data loading and recomputation in a distributed manner.
We also observe that both methods used shorter time to do the recomputation, compared with the normal execution time (about 46ms) before and after iteration 6.
The reason may be the reduced competition in network bandwidth where only the recovering threads need to interact with DSM and enter barrier, while the others have already finished the iteration and been waiting at the barrier asynchronously.

\section{Related Work} \label{relwork}

Most existing works~\cite{mapreduce,spark,storm,flink,flumejava,epic,dryad} focused on developing general-purpose distributed systems for efficient Big Data analytics.
They provide functional primitives to offer great ease-of-use to developers with the compromise of efficiency and flexibility.
In contrast, \bird provides flexible interfaces that allow users to take fine-grained control over distributed threads in a simple yet effective way. 
Recently Husky~\cite{husky} was introduced as a flexible computing framework.
Similar to \bird, it allows objected-oriented programming for distributed workloads.
However, the objects in \bird and Husky are used in different ways. 
Husky adopts object-centric programming model where the data are viewed as objects and objects can communicate with each other via in-class methods. 
In \bird, we regard thread as computation unit that is able to manipulate objects (or other kinds of data) and communicates with other threads via distributed shared memory (based on key-value stores). Moreover, we propose distributed shared data manipulation interfaces to simplify shared data operations and provide an abstraction stack separating user programs from specific key-value store implementation. \bird is also evaluated to be more efficient than Husky over data in large size.

To achieve better performance, many efforts have been devoted to developing specialized systems for particular classes of applications such as graph analytics~\cite{pregel,graphlab,powergraph,trinity, giraph,gps} and machine learning tasks~\cite{petuum,mahout,singa}.
Pregel~\cite{pregel} follows the Bulk Synchronous Parallel (BSP) model and proposes a vertex-centric computation model which is more efficient than MapReduce-based frameworks in distributed graph processing. 
GraphLab~\cite{graphlab} provides asynchronous graph computation to get further performance improvement.
Petuum~\cite{petuum} emerges as a distributed platform for ML applications. It uses parameter server to store intermediate results in the form of matrices. It also introduces Stale Synchronous Parallel (SSP) to trade off between fully synchronous and fully asynchronous modes for model training.
Our work focuses on developing general-purpose distributed systems to cope with complex data analytics pipeline. \bird offers high flexibility to express different classes of applications effectively. We also experimentally show the high efficiency of \bird compared with the specialized systems. 

Prior DSM designs~\cite{ivy} provided strong consistency like sequence consistency, which incurs high communication cost for applications with frequent writes.
Recent researches~\cite{chapel,nelson2015latency} adopted Partitioned Global Address Space (PGAS) model to exploit data locality, where each partition of the global address space is local to a node. 
Different from existing DSM solutions, \bird leverages distributed key-value stores~\cite{memcached,hyperdex,mica,dynamo} to maintain globally shared data. Different key-value store implementation provides slightly different interfaces and functionalities. \bird has decoupled the specific key-value store implementation from shared memory management by introducing a DSM internal layer. We use memcached in our current implementation and \bird can perform a light-weight switch to other key-value stores.

\section{Conclusion}\label{sec:conclusion}

In this paper, we proposed a distributed framework named \bird, towards flexible and efficient data analytics.
It enables users to perform fined-grained control over distributed multi-threading and apply various application-specific optimizations effectively. 
Pthreads-like interfaces for cluster and thread management are offered to simplify programming with \bird. 
\bird leverages the off-the-shelf key-value stores to keep globally shared data. It contains distributed shared data manipulation interfaces so that operations on shared data can be expressed in a similar way as operating local variables. 
We also develop abstraction layers to separate user programs with specific key-value store implementation.
We evaluate the performance of \bird in real applications, and showed that the system is both flexible and efficient.

\newpage


\balance


\bibliographystyle{abbrv}
\bibliography{vldb_sample}  



%
%
%
%

\end{document}